\documentclass[10pt]{article}

\usepackage[a4paper,margin=2cm]{geometry}
\usepackage{amsmath,amsfonts,mathtools,wasysym,stmaryrd,siunitx,cancel,microtype}
\usepackage{graphicx,psfrag,float,overpic,pgfplots,pstool,caption,framed,bchart}
\usetikzlibrary{arrows,shapes,positioning,patterns}
\usepackage{rotating,lmodern,color,multirow,multicol,soul,todonotes,makecell,placeins}
\usepackage[normalem]{ulem}
\usepackage{nomencl}
\usepackage{lineno,hyperref,cleveref}

\pgfplotsset{compat=1.18}

\crefname{equation}{Eq.}{Eqs.}
\hypersetup{pdftitle={Anisotropic Quasi-Brittle Damage Framework Based on the Endurance Surface},pdfauthor={K. Feike, P. Kurzeja, K. Langenfeld and J. Mosler}}

\makenomenclature
\renewcommand*\nompreamble{\begin{multicols}{2}}
\renewcommand*\nompostamble{\end{multicols}}

\newcommand{\B}[1]{\boldsymbol{#1}}
\newcommand{\mc}[1]{\mathcal{#1}}

\newcommand{\of}[1]{\!\left({#1}\right)}

\newcommand{\commentout}[1]{}
\newcolumntype{L}[1]{>{\raggedright\arraybackslash}p{#1}}
\newcolumntype{C}[1]{>{\centering\arraybackslash}p{#1}}
\newcolumntype{R}[1]{>{\raggedleft\arraybackslash}p{#1}}
\newcommand{\el}{^{\text{e}}}
\newcommand{\pl}{^{\text{p}}}
\newcommand{\gr}{_{\varphi}}
\newcommand{\enh}{_\text{enh}}

\newcommand{\mr}[1]{\mathrm{#1}}

\newcommand{\ve}{\varepsilon}
\newcommand{\mac}[1]{\left\langle{#1}\right\rangle}

\newcommand{\partDer}[2]{\dfrac{\partial #1}{\partial #2}}
\newcommand{\totDer}[2]{\dfrac{\mr{d} #1}{\mr{d} #2}}

\newcommand{\trace}[1]{\mr{tr} \left( #1 \right)}

\renewcommand{\exp}[1]{\mr{exp} \left( #1 \right)}

\newcommand{\cccon}{%
	\hspace{0.5ex} \raisebox{0.4ex}{$\cdot$}\hspace{-0.1ex}\raisebox{-0.4ex}{$\cdot$}\hspace{-0.1ex}\raisebox{0.4ex}{$\cdot$} \hspace{0.5ex}%
}

\newtheorem{remark}{Remark}

\renewenvironment{abstract}{%
	\par\bigskip
	\begin{center}
		{\bfseries \abstractname}
	\end{center}
	\noindent\ignorespaces
	\setlength{\parindent}{0pt}%
	\noindent\ignorespaces}{%
	\par\bigskip
}

\newcounter{subfigx}
\newcounter{subfigxabs}
\renewcommand{\thesubfigx}{\alph{subfigx}}

\newsavebox{\subfigbox}

\newcommand{\subfigreset}{%
	\setcounter{subfigx}{0}%
}

\newcommand{\subbfigure}[2][]{%
	\begingroup
	\sbox{\subfigbox}{#2}%
	\begin{minipage}[t]{\wd\subfigbox}%
		\centering
		\usebox{\subfigbox}%
		\par\smallskip
		\stepcounter{subfigxabs}%
		\refstepcounter{subfigx}%
		\parbox[t]{\wd\subfigbox}{%
			\small
			\textbf{(\thesubfigx)}%
			\if\relax\detokenize{#1}\relax
			\else
			\quad #1%
			\fi
			\par
		}%
	\end{minipage}%
	\endgroup
}

\definecolor{TRRRed}{RGB}{222,0,0}
\definecolor{TRRBlue}{RGB}{0, 104, 178}
\definecolor{TRROrange}{RGB}{250, 174, 40}
\definecolor{TRRGreen}{RGB}{135, 192, 33}
\definecolor{TRRPurple}{RGB}{148,0,211}
\definecolor{TRRCyan}{RGB}{0, 200, 255}
\definecolor{TRRGray}{RGB}{215, 215, 215}
\definecolor{TRRWhite}{RGB}{255, 255, 255}
\definecolor{CS1}{RGB}{0.0, 0.0, 162}
\definecolor{CS2}{RGB}{188, 39, 45}

\begin{document}
	
\title{A Unified Gradient-Enhanced Framework for Anisotropic Quasi-Brittle Damage Based on the Endurance Surface in Energy Release Rate Space}

\author{
	K. Feike \quad
	P. Kurzeja \quad
	K. Langenfeld \quad
	J. Mosler\thanks{Corresponding author: \texttt{joern.mosler@tu-dortmund.de}}\\[0.75em]
	\small TU Dortmund University, Institute of Mechanics,\\
	\small Leonhard-Euler-Str. 5, D-44227 Dortmund, Germany
}
\date{}
\maketitle

\begin{abstract}
	This work proposes a novel continuum damage framework for quasi-brittle damage that is based on the endurance-surface concept and uses the energy-release rate as the driving force. Damage evolution is specifically governed by the distance of the thermodynamic driving force from the endurance surface. In contrast to classic failure surfaces, this allows damage to accumulate under both monotonic and cyclic loads even for constant-amplitude loading. The endurance surface directly relates to the physical endurance limit of the material. To obtain mesh-objective results, the formulation is regularized by a micromorphic gradient enhancement. The incorporation of anisotropic damage evolution and the microcrack-closure-reopening effect extends the framework to multiaxial and loading-path-dependent degradation. The chosen prototype damage evolution fulfills three requirements: reasonable physics, computational robustness, and calibration flexibility.
	
	The model is successfully calibrated to both the monotonic response of plain concrete and to the high-cycle fatigue behavior of low-alloy steel. The numerical examples cover monotonic failure of an L-shaped concrete specimen, stress--life behavior under cyclic loading, and combined axial--torsional fatigue. These cases demonstrate how the proposed formulation applies to practical scenarios ranging from standard quasi-brittle fracture benchmarks to classical fatigue characterization and complex multiaxial damage evolution. The examples demonstrate that the formulation captures both progressive monotonic and cyclic degradation. It particularly reproduces characteristic stress--life behavior. The influence of anisotropic degradation becomes especially relevant under multiaxial loading conditions during the near-failure phase.
\end{abstract}


\section{Introduction}\label{sec:introduction}
	Brittle damage plays a central role in the assessment of structural reliability. From a fracture mechanics perspective, it is associated with progressive stiffness degradation, strain localization, stress softening and the formation of crack-like damage patterns. Typical examples include structural elements in buildings and rotating machinery. Quasi-brittle damage can be observed for a wide range of loading conditions and materials, including, for example, cracking in concrete under monotonic loading and cracks in steel induced by high-cycle fatigue~\cite{metal_fatigue_in_engineering_2000}. All of these cases share the common characteristic that the macroscopic response remains predominantly elastic up to failure and damage is driven by the elastic energy release rate \cite{Griffith1921}.
	
	Although monotonic and cyclic quasi-brittle damage evolution indeed share many similarities, the latter poses additional challenges. High-cycle fatigue, for instance, requires a constitutive framework that can distinguish nondamaging cyclic states from states that induce progressive distributed degradation in a predominantly elastic regime. This challenge becomes even more pronounced for multiaxial cyclic loading, because an appropriate anisotropic endurance limit must be elaborated.  Several strategies have been proposed to address this challenge. Crack-growth-based approaches, such as Paris-law formulations, typically focus on the propagation of preexisting cracks by means of sharp interfaces. However, these formulations are primarily based on heuristics rather than thermodynamic principles and distributed damage accumulation as well as crack initiation are not captured~\cite{pugno_generalized_2006,bazant_theory_2014,kirane_size_2016,lo_phase-field_2019}. Alternatively, diffuse representations of cracks, as in phase-field formulations, involve arbitrary crack initiation and propagation. Such formulations have recently been extended to fatigue problems by introducing fatigue-dependent degradation laws~\cite{lo_phase-field_2019,boldrini_non-isothermal_2016,alessi_phenomenological_2018,golahmar_phase_2023}. 
	Nevertheless, incorporating fatigue effects into such frameworks often requires additional phenomenological assumptions and may not directly reflect the thermodynamic structure of continuum damage models. For instance, typical failure surfaces evolve with damage and would therefore require increasing load amplitudes to enable damage accumulation under cyclic loading. Continuum damage mechanics frameworks provide another powerful option to numerically describe distributed material degradation through internal variables~\cite{kachanov1958,lemaitre:hal-03609806}. However, classic formulations were primarily developed for monotonic loading or low-cycle fatigue where inelastic strains dominate. An extension of the continuum damage mechanics framework to high-cycle fatigue is thus required to capture the freely evolving damage field over a large number of cycles~\cite{murakami_metal_2002,EngDamageMech05}.
	
	This work proposes a continuum damage mechanics framework for monotonic and cyclic quasi-brittle damage evolution based on a suitable failure criterion. It must separate load states that lead to progressive damage from those that do not. Inspired by high-cycle fatigue, an endurance surface can define such a criterion. However, existing continuum damage models based on this concept have focused on isotropic formulations. For multiaxial loading, however, an anisotropic generalization is required~\cite{papadopoulos_comparative_1997,papadopoulos_critical_1998}. States outside the endurance surface are admissible and induce damage accumulation even under cyclic loading with constant amplitudes. Embedding this concept into a continuum damage framework provides a promising route to link classic fatigue criteria with thermodynamically consistent material modeling~\cite{ottosen_continuum_2008}. Although the endurance surface concept is motivated by high-cycle fatigue, it can also be applied to monotonic loading as will be shown in the present work.
	
	This work reformulates the endurance-surface concept within a continuum damage mechanics framework. In contrast to classic formulations defined in stress space, the endurance criterion is expressed in terms of the thermodynamic driving force associated with damage. This approach allows a natural integration into the continuum damage modeling structure and facilitates the description of anisotropic degradation through an integrity tensor representation~\cite{STEINMANN19981793}. To obtain physically meaningful and mesh-objective results, the proposed model is moreover combined with a micromorphic gradient enhancement. This approach introduces an intrinsic length scale and controls the spatial distribution of damage~\cite{de_vree_comparison_1995,peerlings_gradient_1996,forest_micromorphic_2009}.
	
	The main contributions of this work are summarized as follows:
	\begin{itemize}
		\item a novel anisotropic endurance-surface framework
		\begin{itemize}
			\item that relates to the physical endurance limit of the material,
			\item based on the energy-release-rate tensor as driving force,
		\end{itemize}
		\item a novel anisotropic quasi-brittle damage evolution suitable for monotonic and cyclic loading (HCF)
		\begin{itemize}
			\item based on the thermodynamically consistent Generalized Standard Materials approach,
			\item integrating reasonable physics, computational robustness, and calibration flexibility,
		\end{itemize}
		\item regularization by means of a tensor-valued micromorphic gradient enhancement,
		\item validation under monotonic loading of plain concrete and cyclic loading of low-alloy steel.
	\end{itemize}
	
	The following section~\ref{sec:fundamentals} introduces the fundamental relations, constitutive equations, and the gradient enhancement. Thereafter, the endurance-surface concept is introduced in section~\ref{sec:methodology}. The core approach is demonstrated, followed by the proposed formulation in terms of thermodynamic driving forces, the damage evolution law, a prototype model, and the incorporation of the gradient enhancement into this prototype model. The latter is thoroughly investigated by means of numerical experiments, addressing both monotonic loads and cyclic loads. Finally, numerical examples in section~\ref{sec:numericalresults} include successful calibration to plain concrete for monotonic loading and to low-alloy steel for cyclic loading. The cyclic loading of low-alloy steel is then further explored by introducing multiaxial loading at finite strains, anisotropic degradation, and the microcrack-closure-reopening effect.
	

\section{Fundamentals}\label{sec:fundamentals}
	This section introduces an established constitutive framework for anisotropic material degradation, cf.~\cite{STEINMANN19981793,Menzel02, Ekh03}. In addition, the fundamentals of the micromorphic gradient enhancement are briefly outlined~\cite{forest_micromorphic_2009,dimitrijevic_method_2008}. As quasi-brittle damage occurs at stress amplitudes well below the macroscopic plastic yield stress of the material, plastic material behavior is therefore neglected in the following section.
	\subsection{Initial energetic framework for local anisotropic damage}\label{ssec:framework}
		\newcommand{\ieps}{\B{b} : \B{\ve}_+ + \B{I}:\B{\ve}_-}
		\newcommand{\iieps}{\B{b} : \left[ \B{\ve}_+ \cdot \B{b} \cdot \B{\ve}_+\right] + \B{\ve}_-:\B{\ve}_- }
		Let the body $\mathcal{B}_0$ be defined in the reference configuration through the material points $\B{X}$, and let $\B{u}\of{\B{X}}$ denote the displacement field. Quasi-brittle damage is typically associated with low strain amplitudes, which motivates the use of a geometrically linearized setting. Accordingly, the strain tensor is defined as
		\begin{align}\label{eq:lin_strain}
			\B{\ve} &= \frac{1}{2}\left[\nabla_{\!\B{X}} \B{u}\of{\B{X}} + \left[\nabla_{\!\B{X}} \B{u}\of{\B{X}}\right]^\text{t} \right]\, \text{.}
		\end{align}
		$\nabla_{\!\B{X}}$ denotes the gradient operator with respect to the reference configuration. A finite-strain extension is employed for the final numerical example, see appendix~\ref{ssec:finite_element_implementation}.
		
		Quasi-brittle damage occurs at stress amplitudes well below the macroscopic plastic yield stress of the material. Plastic material behavior is therefore neglected. Following the works of~\cite{Menzel02, Ekh03}, the Helmholtz energy density $\psi$ is defined for the present quasi-brittle scope as
		\begin{align}\label{eq:helmholtz}
			\psi = \dfrac{\lambda}{2} \, \left[ \B{b} : \B{\ve} \right]^2 + \mu \, \B{b} : \left[ \B{\ve} \cdot \B{b} \cdot \B{\ve} \right]\,\text{.}
		\end{align}
		The operator $:$ denotes the standard double contraction. The symbols $\lambda$ and $\mu$ represent the usual \text{Lam\'{e}} parameters. The second-order tensor $\B{b}$ is referred to as the integrity tensor and is the primary variable of interest. It encodes material degradation analogously to the more commonly used $(1-\textit{damage})$ formulation as follows:
		\begin{alignat}{2}
			\text{integrity}\,\,\B{b} \,\, &\rightarrow \begin{cases*}
				\B{I}, &\text{no damage,}\\
				0 < \text{at least one eigenvalue} < 1, &\text{partially damaged,}\\
				\text{at least one eigenvalue} = 0, &\text{fully damaged.}
			\end{cases*}
		\end{alignat}
	
		The thermodynamic driving forces are obtained as gradients of the Helmholtz energy density~\eqref{eq:helmholtz}
		\begin{align}
			\B{\sigma} &= \partDer{\psi}{\B{\ve}} = \lambda \, \left[\B{b} : \B{\ve}\right] \, \B{b} + 2\,\mu\,\B{b} \cdot \B{\ve} \cdot \B{b}, \quad
			\B{\beta} = -\partDer{\psi}{\B{b}} = - \lambda \, \left[\B{b} : \B{\ve} \right] \, \B{\ve} - 2\,\mu \, \B{\ve} \cdot \B{b} \cdot \B{\ve} \label{eq:beta0}
		\end{align}
		where $\B{\sigma}$ denotes the stress tensor and $\B{\beta}$ is the energy-release-rate tensor, which is associated with the elastic energy. Accordingly, the reduced dissipation inequality of the quasi-brittle case reads
		\begin{align}\label{eq:d_red_plasti}
			\mc{D}^\text{red}_\text{mech} &= \B{\sigma} : \dot{\B{\ve}} - \dot{\psi} = \B{\beta} : \dot{\B{b}} \geq 0 \, \text{.}
		\end{align}
	
		The microcrack-closure-reopening (MCR) effect is implemented following the approach of~\cite{Ekh03} and later investigated as a separate option to account for tension--compression asymmetry. To this end, the strain tensor is additively decomposed into a positive (tensile) component $\B{\ve}_{+}$ and a negative (compressive) component $\B{\ve}_{-}$ using a spectral decomposition with Heaviside function $\mathcal{H}\of{\ve_i}$,
		\begin{align}\label{eq:mcr_decomp}
			\B{\ve} = \B{\ve}_{+} + \B{\ve}_{-} \, \text{,} \quad \B{\ve}_{+} = \sum_{i=1}^{3} \mathcal{H}\of{\ve_{i}} \, \ve_{i} \, \B{N}_{i}^\ve \otimes \B{N}_{i}^\ve \, \text{.}
		\end{align}
		The modified energy density and the associated driving forces read
		\begin{align}
			\psi &= \phantom{-}\dfrac{\lambda}{2} \, \left[\ieps\right]^2 + \mu \, \left[\iieps\right]\label{eq:helmholt_mcr}\\
			\B{\sigma} &=\phantom{-}\partDer{\psi}{\B{\ve}} = \lambda \, \left[\ieps\right] \, \left[ \B{b} : \partDer{\B{\ve}_+}{\B{\ve}} +
			\B{I} : \partDer{\B{\ve}_-}{\B{\ve}}\right] + 2\,\mu\, \left[ \left[ \B{b} \cdot \B{\ve}_+ \cdot \B{b} \right] : \partDer{\B{\ve}_+}{\B{\ve}} +
			\B{\ve}_- : \partDer{\B{\ve}_-}{\B{\ve}} \right] \, \text{,} \label{eq:stress_mcr}
			\\
			\B{\beta} & = {-}\partDer{\psi}{\B{b}} = - \lambda \, \left[ \B{b} : \B{\ve}_+ + {\B{I} : \B{\ve}_-} \, \right] \B{\ve}_+ -
			2 \, \mu \, \B{\ve}_+ \cdot \B{b} \cdot \B{\ve}_+ \, , \label{eq:energyreleaseratefull_mcr} \\
			\widetilde{\B{\beta}} &= \B{\beta} + \lambda \left[ \B{I} : \B{\ve}_- \right] \, \B{\ve}_+ = - \lambda \, \left[ \B{b} : \B{\ve}_+ \right] \B{\ve}_+ -
			2 \, \mu \, \B{\ve}_+ \cdot \B{b} \cdot \B{\ve}_+ \, \text{.} \label{eq:energyreleaseratereduced_mcr}
		\end{align}
		The energy-release-rate tensor~\eqref{eq:energyreleaseratefull_mcr} is analogous to Eq.~\eqref{eq:beta0}\textsubscript{(2)} and the mixed term $\left[\B{I}:\B{\ve}_- \right] \, \B{\ve}_+$ is disregarded in Eq.~\eqref{eq:energyreleaseratereduced_mcr}, following~\cite{Ekh03}, in order to relate damage evolution primarily with tensile strains. Further details of the finite element implementation and the original formulation of the model are provided in appendix~\ref{ssec:finite_element_implementation} and \ref{ssec:framework:o}.
	\subsection{Micromorphic extension}\label{ssec:gradient}
		Numerical damage modeling requires regularization because damage-induced softening otherwise leads to localization and pathological mesh dependence. Among the various gradient-based regularization approaches~\cite{forest_micromorphic_2009,dimitrijevic_method_2008,bazant_crack_1983,rousselier_ductile_1987,dias_da_silva_simple_2004}, a micromorphic approach is adopted~\cite{langenfeld_micromorphic_2020}. In contrast to directly employing the gradient of $\B{b}$, an auxiliary field $\B{\varphi}\of{\B{X}}$ is introduced and energetically coupled to the integrity tensor $\B{b}$ by means of a penalty approach. By doing so, only $C^0$-continuity is required and no inequalities occur at the global level. The enhanced Helmholtz energy density reads
		\begin{align}\label{eq:helmholtz_enh}
			\psi\enh &= \psi\of{ \B{\ve}, \B{b}} + \psi\gr\of{\B{b}, \B{\varphi}, \nabla \B{\varphi}}\,\text{,} \qquad \psi\gr = \frac{c_\mr{b}}{2} \, \left[ \B{\varphi} - \B{b} \right] : \left[ \B{\varphi} - \B{b}\right] + \frac{c_\mr{b} \, \ell_\mr{b}^2}{2} \, \nabla \B{\varphi} \cccon \nabla \B{\varphi}
		\end{align}
		with penalty parameter $c_\mr{b}$ and length-scale parameter $\ell_\mr{b}$, which is related but not identical to the localization width. The operator $\cccon$ denotes a triple contraction, i.e., $\B{A} \cccon \B{A} = A_{ijk} A_{ijk}$ in Einstein notation. The additional thermodynamic driving forces are given by
		\begin{align}\label{eq:duals_enh}
			\B{\omega} = \partDer{\psi\enh}{\B{\varphi}} = c_\mr{b} \, \left[\B{\varphi} - \B{b}\right]\,\text{,} \qquad
			\B{\Omega} = \partDer{\psi\enh}{\nabla \B{\varphi}} = c_\mr{b} \, \ell_\mr{b}^2 \, \nabla \B{\varphi}\,\text{,} \qquad 
			\B{\beta}\enh = -\partDer{\psi\enh}{\B{b}} = \B{\beta} + \B{\omega} \, \text{.}
		\end{align}
		Assuming that microforces $\B{\omega}$ and microstresses $\B{\Omega}$ are purely energetic, cf.~\cite{langenfeld_curvature_2023}, the stress power reads ${\mathcal{P}=\B{\sigma}:\dot{\B{\ve}}+\B{\omega}:\dot{\B{\varphi}}+\B{\Omega}\cccon\nabla \dot{\B{\varphi}}}$, and the reduced dissipation inequality exhibits the same structure as the local model (Eq.~\eqref{eq:d_red_plasti}). For a sufficiently large coupling parameter $c_\mr{b}$ ($\B{\varphi} \rightarrow \B{b}$), the reduced dissipation inequality~\eqref{eq:d_red_plasti} is recovered as the limiting case, cf.~\cite{langenfeld_micromorphic_2020}.
		
		In this manuscript, the micromorphic enhancement is used as a numerically efficient approximation for a standard gradient continuum. Accordingly, the penalty parameter is chosen sufficiently large. The length parameter $\ell_\mathrm{b}$ is moreover chosen to be sufficiently small compared to the structural diameter and sufficiently large compared to the underlying finite element mesh for each numerical example. Once more it is emphasized that although $\ell_\mathrm{b}$ implicitly defines the width of the localization zone ($\ell_\mathrm{lz}$), the two lengths are not identical (i.e., $\ell_\mathrm{lz}\neq \ell_\mathrm{b}$). Moreover, in line with phase-field theory, homogeneous Neumann boundary conditions have been chosen for the introduced auxiliary field.

\section{Using the endurance surface for a novel quasi-brittle anisotropic damage framework}\label{sec:methodology}
	\subsection{The endurance-surface concept}\label{ssec:endurance_surf}
\begin{figure}[ht]
			\centering
			\subfigreset
			\psfrag{t}[c][r]{$t$ [-]}
			\psfrag{b1}[c][c]{\small energy-release rate eigenvalue $\bar{\beta}_1$ [-]}
			\psfrag{b2}[c][c]{\small energy-release rate eigenvalue $\bar{\beta}_2$ [-]}
			\psfrag{fs}[c][r]{~~\underline{failure surface}}
			\psfrag{es}[c][l]{\underline{endurance surface}}
			\psfrag{tf1}[c][l]{\rotatebox{15}{\makebox{~~\textcolor{CS1}{\small elastic loading}}}}
			\psfrag{tf2}[c][c]{\rotatebox{15}{\colorbox{white}{\textcolor{TRRRed}{\small damage}}}}
			\psfrag{tf3}[c][l]{\rotatebox{-30}{\makebox{\textcolor{CS1}{\small elastic}}}}
			\psfrag{expl}[l][l]{}
	 		\psfrag{expl2}[c][c]{\scriptsize
			\begin{tabular}{l}
				current\\[1.5ex]
				initial
	 		\end{tabular}}
			\psfrag{ts1}[c][l]{\rotatebox{15}{\makebox{\textcolor{CS1}{\small elastic loading}}}}
			\psfrag{ts2}[c][r]{\rotatebox{-35}{\textcolor{TRRRed}{\small damage}}}
			\psfrag{ts3}[c][l]{\rotatebox{-30}{\textcolor{CS1}{\small elastic}}}
			\subbfigure[Classic failure-surface formulation: cyclic loading response. The surface evolves with damage, so damage accumulates only during the first cycle.]{%
				\includegraphics[width=0.45\textwidth]{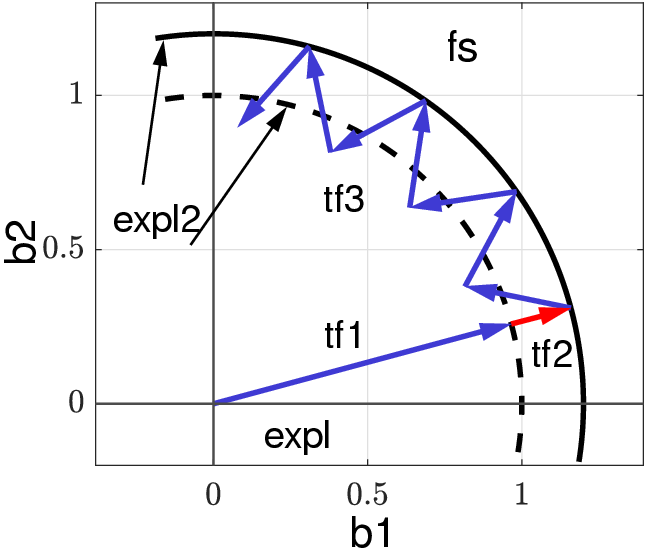}
			}
			\hfill
			\subbfigure[Endurance-surface formulation: cyclic loading response. The surface does not need to evolve and damage accumulates over all cycles.]{%
				\includegraphics[width=0.45\textwidth]{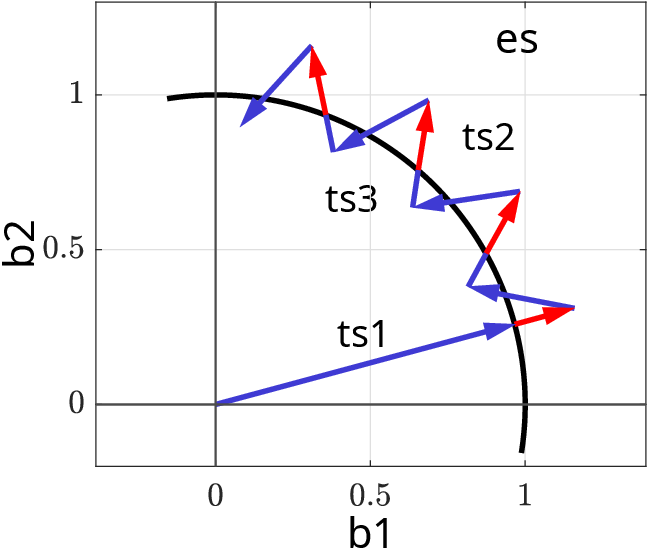}
			}
			\caption{Endurance-surface concept: Comparison of cyclic loading behavior between classic failure surface formulations and the endurance-surface formulation -- red: damage evolution, blue: purely elastic.}\label{fig:enduranceadvantage}
		\end{figure}

	 	The proposed framework is based on the concept of an endurance surface~\cite{ottosen_continuum_2008,holopainen_continuum_2016,ottosen_enhanced_2018,lindstrom_continuous-time_2020,tveit_continuum_2024}, which remedies the damage stagnation associated with classic failure surfaces under cyclic loading. The key difference between the endurance surface and a failure surface is that states outside the endurance surface are admissible and lead to damage evolution -- similar to viscoplastic models governed by an overstress. Typical failure surfaces evolve with damage and would therefore require increasing load amplitudes to enable damage accumulation under cyclic loading (Fig.~\ref{fig:enduranceadvantage}~(a)). In contrast, the endurance surface allows cyclic damage accumulation, even under loading at constant amplitudes (Fig.~\ref{fig:enduranceadvantage}~(b)).
		
		Quantitatively, damage evolution is related to the distance between the current state and the endurance surface. The evolution is further restricted by two conditions: First, only states outside the endurance surface induce damage evolution. Second, damage evolves only for positive loading rates. The latter ensures that no damage develops during unloading.
		
		In the spirit of the nonassociative potential used in the original formulation for ductile damage, cf.~\cite{Ekh03}, the endurance surface is defined in the energy space as
		\begin{align}\label{eq:endurance_surface}
			f\of{\widetilde{\B{\beta}}, \B{b}} &= \sqrt{\frac{\eta_\text{i}}{2} \, \left[ \B{b}^{r} : \widetilde{\B{\beta}} \right]^2 + 		\frac{\eta_\text{a}}{2} \, \B{b}^{r} : \left[ \widetilde{\B{\beta}} \cdot \B{b}^{r} \cdot \widetilde{\B{\beta}} \right]} - 1.
		\end{align}
		The first two mixed invariants of the modified energy-release-rate tensor $\widetilde{\B{\beta}}$ enter in combination with the integrity tensor $\B{b}$. Extending~\cite{ottosen_continuum_2008}, they allow for an anisotropic description of degradation. Furthermore, and in contrast to~\cite{ottosen_continuum_2008}, the endurance surface is defined within the natural space of the energy-release-rate tensor. Model parameters $\eta_\text{i}$ and $\eta_\text{a}$ control the contributions of isotropic and anisotropic damage evolution, respectively. Exponent $r$ allows to control expansion or contraction of the endurance surface. If $\eta_\text{i}$ and $\eta_\text{a}$ are nonnegative, Eq.~\eqref{eq:endurance_surface} is homogeneous of degree one. In general, the endurance surface can be controlled by $\eta_\mr{i}$ to remain constant, expand (delaying damage), or shrink (accelerating damage). In the anisotropic case ($\eta_\mr{a} \neq 0$), the endurance surface may additionally change its shape, allowing directional control over the severity of damage evolution.
		
		\subsection{Damage evolution within the endurance-surface framework}\label{ssec:damage_evol_endsurf}
			The evolution equation for the integrity tensor $\B{b}$, and thus for damage, is formulated in the spirit of generalized standard materials~\cite{Halphen1975SurLM} as
			\begin{align}\label{eq:damage_evol_0}
				\dot{\B{b}} &=\mathcal{H}\of{f} \, \mac{\dot{f}} \, \gamma\of{f} \, 			\partDer{\Gamma\of{\B{\beta}}}{\B{\beta}}\, \text{.}
			\end{align}
			Extending~\cite{ottosen_continuum_2008}, Eq.~\eqref{eq:damage_evol_0} accounts for anisotropic material degradation. $\mathcal{H}\of{f}$ and $\mac{\dot{f}} = \mr{max}\of{\dot{f},0}$ ensure that the integrity tensor $\B{b}$ evolves only when $f \geq 0$ and $\dot{f} > 0$, respectively. The latter plays a role similar to a plastic multiplier by incorporating a pseudo-time component, as known from classic plasticity. The function $\gamma\of{f}$ scales the evolution equation depending on the current state $f\of{\widetilde{\B{\beta}},\B{b}}$ and may be chosen from the class of nonnegative functions in order to fulfill thermodynamic constraints. The potential $\Gamma\of{\B{\beta}}$ controls the evolution direction and must be convex with respect to its argument. Furthermore, it must contain the origin and be nonnegative in order to satisfy thermodynamic requirements (analogous to the generalized standard materials approach~\cite{Halphen1975SurLM}). Moreover, it is advantageous to choose a homogeneous function of degree one in order to decouple the evolution amplitude ($\gamma\of{f}$) from the evolution direction ($\partial_{\B{\beta}} \Gamma\of{\B{\beta}}$). Further details regarding thermodynamic consistency are provided in appendix~\ref{ssec:appendix_a:thermo}. By enforcing nonnegative eigenvalues of tensor $\B{b}$ during tensor exponentiation by a numerically smoothed maximum function, the constraint $\text{eig}\of{\B{b}} \geq 0$ is not violated provided the temporal resolution is chosen sufficiently small. The evolution of $\text{eig}\of{\dot{\B{b}}}$ naturally converges to zero as $\text{eig}\of{\B{b}}$ approaches zero.
	\subsection{Prototype model}
			A specific damage-evolution law must satisfy three requirements: be physically reasonable, ensure computational robustness, and remain flexible enough for calibration to various materials. The previously introduced evolution equation for the integrity tensor $\B{b}$ is hence specified by
			\begin{align}
				\gamma\of{f} &= K_1\,\exp{p_1\,f} + \frac{K_2}{2^{p_2}}\,\left[ \sqrt{4+f^2} + f \right]^{p_2}\label{eq:end_gamma},\\
				\Gamma\of{\B{\beta}} &= \sqrt{\frac{\eta_\text{i}\, \trace{\B{\beta}}^2 + \eta_\text{a}\, 	\trace{\B{\beta}^2}}{\eta_\text{i}+\eta_\text{a}}}.\label{eq:end_Gamma_pot}
			\end{align}
			The parameters $K_1$, $K_2$, $p_1$ and $p_2$ control the scaling function $\gamma\of{f}$. The exponential term follows~\cite{ottosen_continuum_2008}, while the power-law term is added to provide more flexibility during the calibration. Parameters $\eta_\mr{i}$ and $\eta_\mr{a}$ control the normalized isotropic and anisotropic evolution direction through the invariants of $\B{\beta}$. Purely isotropic evolution is obtained by setting $\eta_\text{a}=0$ and initializing $\B{b}$ as a spherical tensor, if desired.
	\subsection{Gradient enhancement of the endurance surface}\label{ssec:endurance_enhanced}
		To incorporate the gradient enhancement, the endurance surface~\eqref{eq:endurance_surface} is modified similarly to~\cite{langenfeld_micromorphic_2020} as follows
		\begin{align}\label{eq:endurance_surface_grf}
			f\enh\of{\widetilde{\B{\beta}},\,\B{\omega},\,\B{b}} &= \sqrt{\frac{\eta_\text{i}}{2} \, \left[ \B{b}^{r} : \widetilde{\B{\beta}} \right]^2 + \frac{\eta_\text{a}}{2} \, \B{b}^{r} : \left[ \widetilde{\B{\beta}} \cdot \B{b}^{r} \cdot \widetilde{\B{\beta}} \right]} - \frac{1}{2} \, \left[ \sqrt{ \epsilon + \left[1 + \frac{\B{b}^{q} : \B{\omega}}{\omega_\mathrm{n}}\right]^2} + 1 + \frac{\B{b}^{q} : \B{\omega}}{\omega_\mathrm{n}} \right] \, \text{.}
		\end{align}
		The enhancement replaces the former ``1-threshold'' and defines a nonnegative and monotonically increasing function of ${\B{b}^q:\B{\omega}}/{\omega_\mathrm{n}}$ with exponent $q$ and predefined normalization parameter $\omega_\mathrm{n} = 1\,\si{\mega\pascal}$ to allow for unambiguous coupling through $c_\mathrm{b}$. This formulation is chosen to achieve a self-stabilizing evolution between the integrity field $\B{b}$ and the auxiliary field $\B{\varphi}$. Accordingly, the rate $\dot{\B{b}}$ increases or decreases toward perfect coupling ($\B{\omega}= c_\mr{b}\left[\B{\varphi}-\B{b}\right]\rightarrow\B{0}$). The coupling parameter $c_\mr{b}$ must be sufficiently large, and the numerical tolerance is set to $\epsilon = 10^{-4}$. 
		
		It is noteworthy that a direct gradient enhancement may lead to reversed, and therefore unphysical, damage evolution, cf. appendix~\ref{ssec:naive_enhancement}.
	\subsection{1D example -- assessment of the gradient enhancement for monotonic loading and cyclic loading}\label{ssec:exp_grad}
		The following methodological examples aim to demonstrate the effect of the gradient enhancement. They investigate the proposed endurance-surface-based damage formulation under strain localization before the next section continues with specific examples for plain concrete and low-alloy steel.
		
\begin{figure}[!ht]
				\centering
				\subfigreset
				\subbfigure[Geometry and loading conditions.]{%
					\psfrag{x}[c][c]{\scalebox{0.9}{\makebox{$x$}}}
					\psfrag{g2}[c][c]{\scalebox{0.9}{\makebox{$1.05\,\eta_\mr{i}$}}}
					\psfrag{L}[c][c]{\scalebox{0.9}{\makebox{$\ell = 100 \si{\milli\meter}$}}}
					\psfrag{A}[l][c]{\scalebox{0.9}{\makebox{$A = 100\, \si{\milli\meter^2}$}}}
					\psfrag{uf}{$u$, $F$}
					\psfrag{f}[c][c]{\scalebox{0.9}{\makebox{force $F \, [\si{\newton}]$}}}
					\psfrag{u}[c][c]{\scalebox{0.9}{\makebox{displacement $u \, [\si{\micro\meter}]$}}}
					\includegraphics[width=0.45\textwidth]{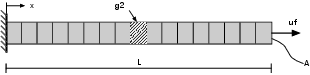}%
				}
				
				\par\bigskip
				\noindent\makebox[\textwidth][c]{\hspace{1.5em}\textbf{monotonic loading}}
				\par\medskip
				
				\subbfigure[Local formulation: force--displacement curves for different mesh resolutions.]{%
					\begin{tikzpicture}
						\node[anchor=south west, inner sep=0] (image) at (0,0)
						{\psfrag{f}[c][c]{\scalebox{0.75}{\makebox{force $F \, [\si{\newton}]$}}}
							\psfrag{u}[c][c]{\scalebox{0.75}{\makebox{displacement $u \, [\si{\micro\meter}]$}}}
							\includegraphics[width=0.285\textwidth]{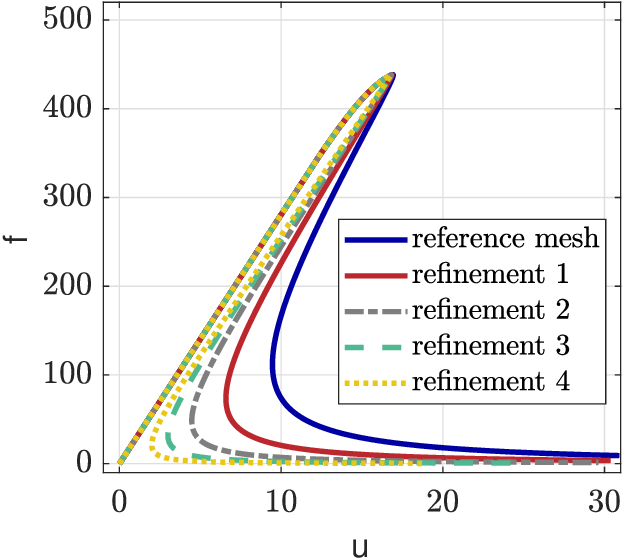}};
						\begin{scope}[x={(image.south east)}, y={(image.north west)}]
							\node[anchor=south west] at (0.25, 0.8) {\scalebox{0.8}{local}};
						\end{scope}
					\end{tikzpicture}
				}
				\hfill
				\subbfigure[Integrity field along the bar for the local (left) and gradient-enhanced (right) formulations at $u=30~\si{\micro\meter}$ for different mesh resolutions.]{%
					\begin{tikzpicture}
						\node[anchor=south west, inner sep=0] (image) at (0,0)
						{\psfrag{x}[c][c]{\scalebox{0.75}{\makebox{normalized axial coordinate $x/\ell$}}}
								\psfrag{b}[c][c]{\scalebox{0.75}{\makebox{integrity value $b_\text{iso} \, [-]$}}}
								\includegraphics[width=0.285\textwidth]{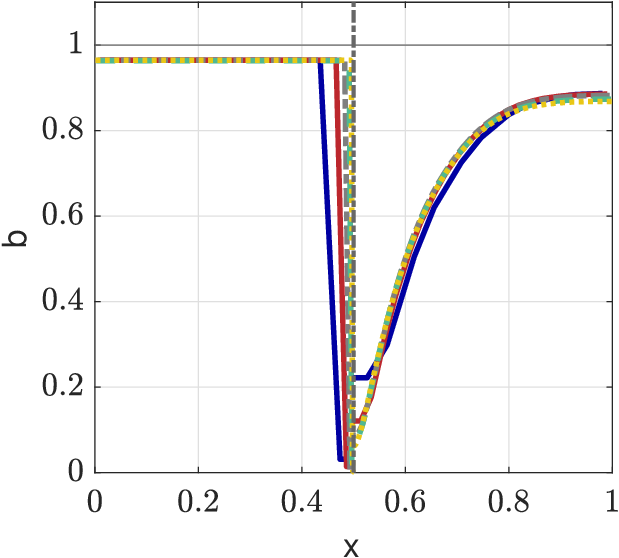}};
						\begin{scope}[x={(image.south east)}, y={(image.north west)}]
								\node[anchor=south west] at (0.25, 0.2) {\scalebox{0.8}{local}};
								\node[anchor=south west] at (0.65, 0.2) {\scalebox{0.8}{enhanced}};
							\end{scope}
					\end{tikzpicture}
				}
				\hfill
				\subbfigure[Gradient-enhanced formulation: force--displacement curves for different mesh resolutions.]{%
					\begin{tikzpicture}
						\node[anchor=south west, inner sep=0] (image) at (0,0)
						{\psfrag{f}[c][c]{\scalebox{0.75}{\makebox{force $F \, [\si{\newton}]$}}}
								\psfrag{u}[c][c]{\scalebox{0.75}{\makebox{displacement $u \, [\si{\micro\meter}]$}}}
								\includegraphics[width=0.285\textwidth]{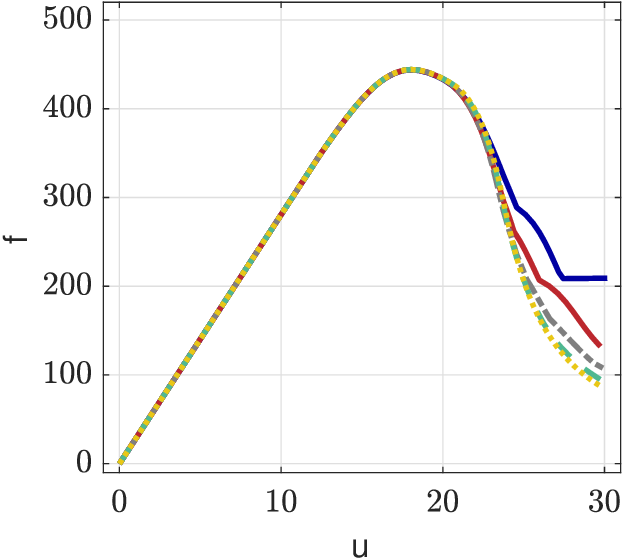}};
						\begin{scope}[x={(image.south east)}, y={(image.north west)}]
								\node[anchor=south west] at (0.25, 0.8) {\scalebox{0.8}{enhanced}};
							\end{scope}
					\end{tikzpicture}
				}
			
				\par\bigskip
				\noindent\makebox[\textwidth][c]{\hspace{1.8em}\textbf{cyclic loading}}
				\par\smallskip
				
				\subbfigure[Local formulation: force--displacement curves during cyclic loading with incremental load increase.]{%
					\begin{tikzpicture}
						\node[anchor=south west, inner sep=0] (image) at (0,0)
						{%
								\psfrag{t1}[c][l]{\tiny\begin{tabular}{l}
												displ.\\
												increase
											\end{tabular}}
								\psfrag{t2}[c][c]{\tiny\textcolor{CS2}{\begin{tabular}{l}
														damage\\
														over\\
														cycles
												\end{tabular}}}
								\psfrag{t3}[c][c]{\tiny\textcolor{CS1}{snap-back}}
								\psfrag{u}[c][c]{\scalebox{0.75}{\makebox{displacement $u \, [\si{\micro\meter}]$}}}
								\psfrag{f}[c][c]{\scalebox{0.75}{\makebox{force $F \, [\si{\newton}]$}}}
								\includegraphics[width=0.285\textwidth]{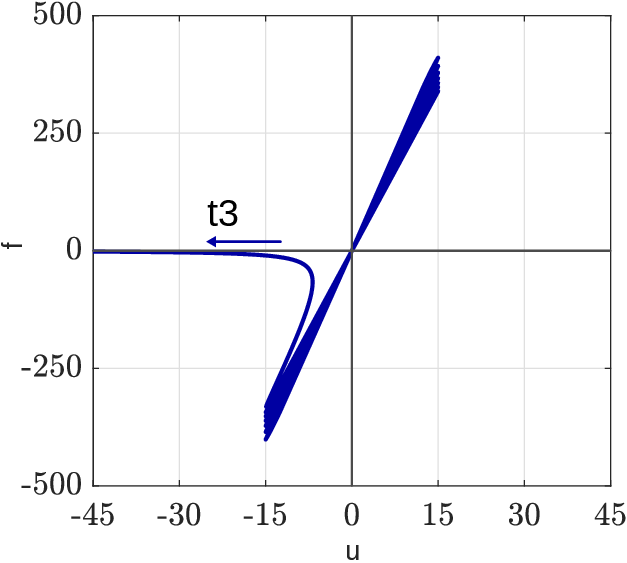}};
						\begin{scope}[x={(image.south east)}, y={(image.north west)}]
								\node[anchor=south west] at (0.75, 0.2) {\scalebox{0.8}{local}};
							\end{scope}
					\end{tikzpicture}
				}
				\hfill
				\subbfigure[Integrity field along the bar at the final loading step for the local (left) and gradient-enhanced (right) formulations.]{%
					\raisebox{-2.3pt}{%
					\begin{tikzpicture}
							\node[anchor=south west, inner sep=0] (image) at (0,0)
							{\psfrag{x}[c][c]{\scalebox{0.75}{\makebox{normalized axial coordinate $x/\ell$}}}
									\psfrag{b}[c][c]{\scalebox{0.75}{\makebox{integrity value $b_\text{iso} \, [-]$}}}
									\includegraphics[width=0.282\textwidth]{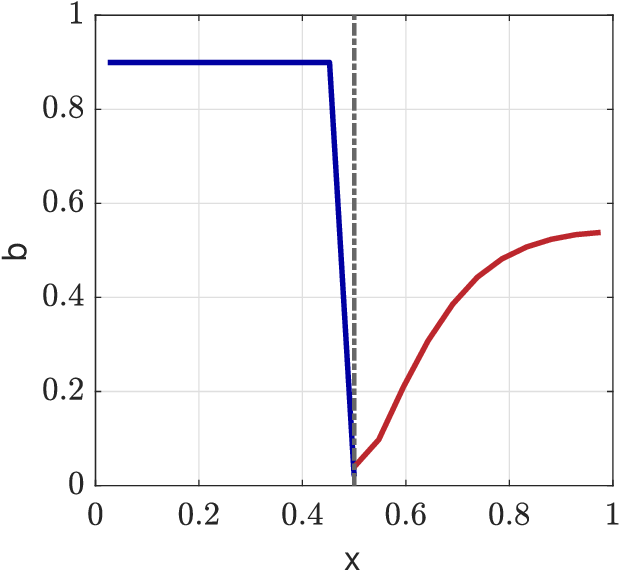}};
							\begin{scope}[x={(image.south east)}, y={(image.north west)}]
									\node[anchor=south west] at (0.25, 0.2) {\scalebox{0.8}{local}};
									\node[anchor=south west] at (0.65, 0.2) {\scalebox{0.8}{enhanced}};
								\end{scope}
						\end{tikzpicture}
				}}
				\hfill
				\subbfigure[Gradient-enhanced formulation: force--displacement curves during cyclic loading with incremental load increase.]{%
					\begin{tikzpicture}
						\node[anchor=south west, inner sep=0] (image) at (0,0)
						{%
								\psfrag{t1}[c][l]{\tiny\begin{tabular}{l}
												displ.\phantom{.}\\
												increase\phantom{.}
											\end{tabular}}
								\psfrag{t2}[c][c]{\tiny\textcolor{CS2}{\begin{tabular}{l}
														\phantom{.}damage\\
														\phantom{.}over\\
														\phantom{.}cycles
													\end{tabular}}}
								\psfrag{t3}[l][l]{\tiny\textcolor{CS1}{snap-back}}
								\psfrag{u}[c][c]{\scalebox{0.75}{\makebox{displacement $u \, [\si{\micro\meter}]$}}}
								\psfrag{f}[c][c]{\scalebox{0.75}{\makebox{force $F \, [\si{\newton}]$}}}
								\includegraphics[width=0.285\textwidth]{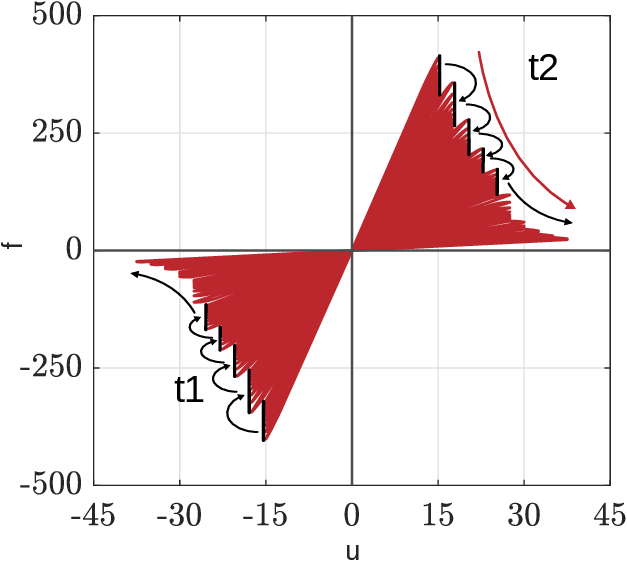}};
						\begin{scope}[x={(image.south east)}, y={(image.north west)}]
								\node[anchor=south west] at (0.6, 0.2) {\scalebox{0.8}{enhanced}};
							\end{scope}
					\end{tikzpicture}
				}
				\caption{Numerical analysis of a 1D bar specimen: monotonic localization behavior (top) and cyclic degradation under incremental load increase (bottom) of plain concrete.}\label{fig:d1rodresults}
			\end{figure}

		A one-dimensional bar of length $\ell = 100\, \si{\milli\meter}$ and cross-sectional area $A=100\, \si{\milli\meter^2}$ is considered, as illustrated in Fig.~\ref{fig:d1rodresults}~(a). One end of the bar is clamped while the other is subjected to loading. An imperfection is introduced in the middle element by increasing its initial damage modulus to $1.05\,\eta_\mr{i}$ to trigger localization. Although the following experiments are primarily illustrative in nature, the model parameters correspond to the parameter set used for plain concrete introduced later in section~\ref{ssec:plain_concrete}. Additional numerical experiments addressing mesh convergence for a 2D compact tension specimen are presented in~\ref{ssec:2d_conv}.
		\paragraph{Monotonic localization behavior}
			To investigate the localization behavior of the proposed damage formulation, the bar is discretized using $\left\{ 11, 21, 41, 81, 161 \right\}$ elements. The specimen is subjected to uniaxial tension using an arc-length method, cf.~\cite{crisfield_fast_1981}. Figures~\ref{fig:d1rodresults}~(b--d) compare the resulting force--displacement responses for the local and gradient-enhanced formulations for different mesh resolutions.
	
			The local formulation exhibits pronounced snap-back behavior after reaching a peak load of $F_\text{max}=430 \, \si{\newton}$ for all mesh refinements (Fig.~\ref{fig:d1rodresults}~(b)). Within the snap-back regime, the area under the force--displacement curve (equivalent to the fracture energy) varies significantly with the element size. This mesh dependence is characteristic of local damage formulations undergoing strain localization. The same mesh dependence is observed for the integrity field (Fig.~\ref{fig:d1rodresults}~(c) left), where damage localizes entirely within the respective imperfect element.
			
			In contrast, the gradient-enhanced formulation exhibits converging softening for all mesh refinements (Fig.~\ref{fig:d1rodresults}~(d)). From the second refinement onward, the force--displacement responses follow nearly identical paths, indicating convergence of the dissipated fracture energy with decreasing element size. Deviations observed for the coarsest meshes result from an insufficient resolution of the localization zone. The integrity fields accordingly spread smoothly over a finite region of the bar and are controlled by the length scale of the gradient enhancement (Fig.~\ref{fig:d1rodresults}~(c) right).
		\paragraph{Cyclic degradation behavior}
			A similar evolution of the integrity fields is observed under cyclic loading with gradually increasing displacement amplitudes from $15$ to $22.5 \, \si{\micro\meter}$, for a discretization of 21 elements and disabled MCR effect (Fig.~\ref{fig:d1rodresults}~(f)). Partial degradation occurs already during the first cycle for both the local and the gradient-enhanced formulations. In the seventh cycle, the compressive branch of the local model, however, enters snap-back behavior, indicating abrupt structural failure (Fig.~\ref{fig:d1rodresults}~(e)). In contrast, the gradient-enhanced model exhibits progressive degradation throughout the entire loading history without transitioning to snap-back behavior (Fig.~\ref{fig:d1rodresults}~(g)). It should be emphasized that this setup uses the material parameters of plain concrete, but the loading conditions were specifically chosen to highlight the difference between the local and the gradient-enhanced model.
			
\begin{figure}[ht]
				\centering
				\subfigreset
				\psfrag{t1}[c][c]{\tiny\begin{tabular}{l}
						original\\
						stiffness\\
						recovered
					\end{tabular}}
				\psfrag{t2}[c][c]{\tiny\textcolor{CS2}{\begin{tabular}{l}
							damage\\
							over\\
							cycles
						\end{tabular}}}
				\psfrag{t3}[l][l]{\tiny\textcolor{CS1}{snap-back}}
				\psfrag{u}[c][c]{\scalebox{0.75}{\makebox{displacement $u \, [\si{\micro\meter}]$}}}
				\psfrag{f}[c][c]{\scalebox{0.75}{\makebox{force $F \, [\si{\newton}]$}}}
				\psfrag{n}[c][c]{\scalebox{0.75}{\makebox{cycles to failure $N \, [-]$}}}
				\psfrag{h}[c][c]{\scalebox{0.75}{\makebox{element size $h \, [\si{\milli\meter}]$}}}
				\subbfigure[Local formulation: force--displacement curve during cyclic loading with incremental load increase.]{
					\includegraphics[width=0.29\textwidth]{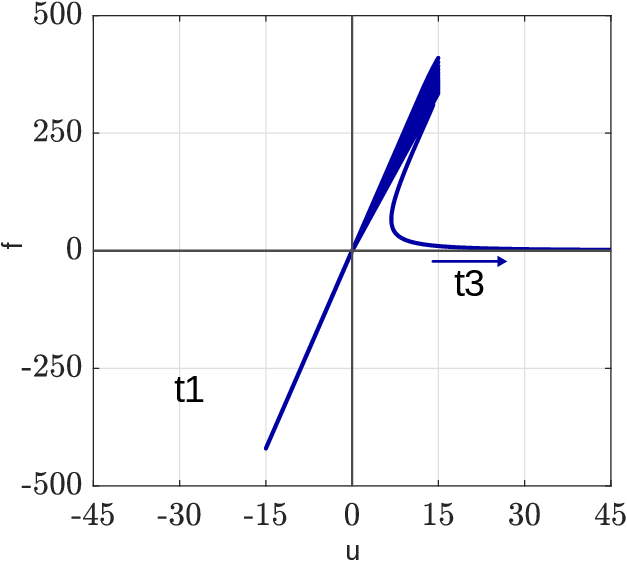}
				}
				\hspace{0.5cm}
				\subbfigure[Gradient-enhanced formulation: force--displacement curve during cyclic loading with incremental load increase.]{
					\includegraphics[width=0.29\textwidth]{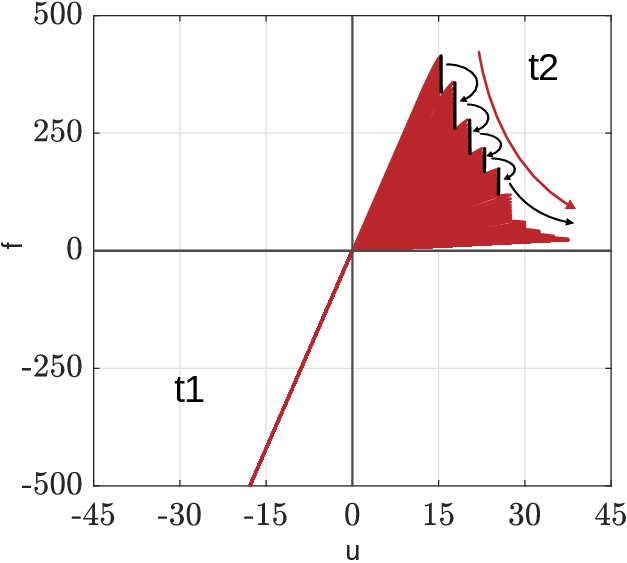}
				}
				\hspace{0.5cm}
				\raisebox{-2.6pt}{%
				\subbfigure[Cycles to failure versus element size for the gradient-enhanced formulations with and without the MCR effect (altered model parameters).]{%
					\includegraphics[width=0.295\textwidth]{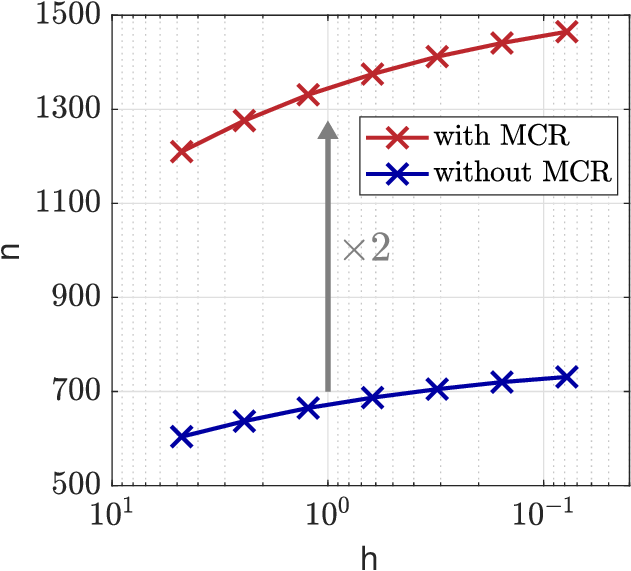}
				}}
				\caption{Numerical analysis of a 1D bar specimen: cyclic degradation with tension--compression asymmetry due to the MCR effect and mesh-convergence analysis.}\label{fig:d1rodresults:cyclic:academic:mcr}
			\end{figure}

		\paragraph{Cyclic degradation with MCR effect}
			The experiment is repeated with the MCR tension--compression decomposition enabled. The numerical smoothing of the Heaviside function in Eq.~\eqref{eq:mcr_decomp} follows~\cite{Ekh03} using the numerical parameters $g_0=0$, $x_0=0$, and $x_\mathrm{R}=10^{-6}$. This allows the influence of the tension--compression asymmetry on the cyclic degradation behavior to be assessed. In this case, the displacement amplitude is increased every twentieth cycle over a total of $200$ cycles. When the MCR decomposition is active (Fig.~\ref{fig:d1rodresults:cyclic:academic:mcr}), the tensile regime behaves similarly to that observed in Figs.~\ref{fig:d1rodresults}~(e) and~(g).
			However, the compressive branches remain elastic due to the tension--compression asymmetry introduced by the MCR effect. In this example, the local model transitions to snap-back behavior during the fifteenth cycle (Fig.~\ref{fig:d1rodresults:cyclic:academic:mcr}~(a)), whereas the enhanced model does not exhibit snap-back behavior throughout the entire loading history (Fig.~\ref{fig:d1rodresults:cyclic:academic:mcr}~(b)). The experiments with the MCR effect enabled produce integrity fields that are virtually identical to those obtained in the experiments without the MCR effect (Fig.~\ref{fig:d1rodresults}~(f)).
			
		\paragraph{Mesh convergence of the gradient enhancement}
			A final experiment investigates the mesh convergence of the gradient-enhanced model in the cyclic regime. Both instances are considered, i.e., simulations without the MCR effect and with the MCR effect. The geometric setup remains identical to the previous experiment. To define a point of failure for cyclic loading without requiring an incremental load increase, the model parameters, and in particular damage exponent $r$, are modified such that degradation accelerates with increasing damage. Consequently, failure is defined as the onset of snap-back. The results are shown in Fig.~\ref{fig:d1rodresults:cyclic:academic:mcr}~(c). Finer mesh resolution increases predicted cycles to failure in both displacement-driven scenarios. Furthermore, the results show that, for the present setup, the number of cycles required to reach a comparable damage state is approximately doubled when the MCR effect is considered. Consequently, the MCR effect can also be implemented by a halved damage accumulation rate governed by coefficients $K_1$ and $K_2$.
			
			Overall, the monotonic and cyclic experiments demonstrate the necessity of regularization for the proposed damage formulation. The gradient enhancement provides mesh-objective structural responses and stable damage evolution under both loading conditions. Moreover, the cyclic simulations illustrate the applicability of the proposed endurance-surface formulation to fatigue problems. In addition, the structural responses in the cyclic setting with the MCR effect exhibit the intended tension--compression asymmetry.

\section{Numerical examples}\label{sec:numericalresults}
	The following numerical examples assess the capabilities of the proposed model under increasing complexity in terms of load conditions, the MCR effect, and geometry. The model is first calibrated for two different materials and loading conditions: monotonic loading of plain concrete and cyclic loading of low-alloy steel. The setup for plain concrete was already used for the investigation of the gradient enhancement above. The calibrated steel setup is moreover validated against additional experimental data. Finally, the axial--torsional loading of a cylindrical steel specimen is considered to mimic practical laboratory conditions, including the highest model complexity: finite strains, the MCR effect, and anisotropic degradation.
	
	\subsection{Plain concrete -- model calibration and assessment against monotonic experiments}\label{ssec:plain_concrete}
\begin{figure}[ht]
	 			\centering
	 			\psfrag{disp}[c][c]{\scalebox{0.9}{\makebox{displacement $u \, [\si{\milli\meter}]$}}}
	 			\psfrag{frc}[c][c]{\scalebox{0.9}{\makebox{force $F \, [\si{\kilo\newton}]$}}}
	 			\psfrag{fu}[l][c]{\scalebox{0.8}{\raisebox{-18pt}{\makebox{$u$,$F$}}}}
	 			\psfrag{l1}[c][c]{\scalebox{0.6}{\makebox{$500\,\si{\milli\meter}$}}}
	 			\psfrag{l2}[c][c]{\scalebox{0.6}{\makebox{$250\,\si{\milli\meter}$}}}
	 			\begin{tikzpicture}
	 				\node[anchor=south west,inner sep=0] (main) at (0,0)
	 				{\psfrag{DA}[l][c]{\scalebox{0.75}{\makebox{$\Bigg \}$
	 								\begin{tabular}{@{}l}
	 									experiments~\cite{winkler_experimental_2001}\\
	 									up to $u=0.4\,\si{\milli\meter}$
 									\end{tabular}}}}
					\psfrag{DD}[l][c]{\scalebox{0.75}{\makebox{$\phantom{\Bigg \}}$model prediction}}}
					\includegraphics[width=0.5\textwidth]{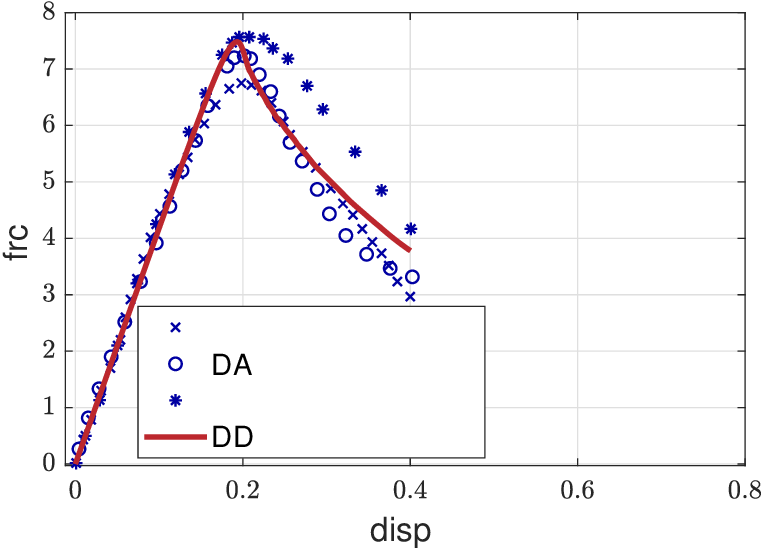}};
					\begin{scope}[x={(main.south east)},y={(main.north west)}]
						\node[anchor=north east] at (0.94,0.99) {\includegraphics[width=0.19\textwidth]{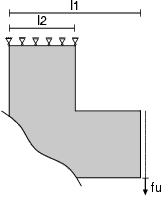}};
					\end{scope}
				\end{tikzpicture}
				\caption{L-shaped specimen made of plain concrete -- model calibration and assessment against monotonic experiments: geometry and boundary conditions of the calibration setup and corresponding force--displacement response. Experimental curves from~\cite{winkler_experimental_2001} are compared with the response of the calibrated model under monotonic loading.}\label{fig:calibration1}
			\end{figure}

		\paragraph{Setup}
	 		The experimental setup, elastic parameters, and force--displacement curves were adopted from the experiments reported in \cite{winkler_experimental_2001} and are sketched in Fig.~\ref{fig:calibration1}. Three force--displacement curves were extracted from these experiments and used as the calibration target. The assumption of plane-strain is used here, cf.~\cite{muixi_combined_2021}.
	 
	 		The specimen geometry shown in Fig.~\ref{fig:calibration1} is discretized using roughly $3000$ quadrilateral finite elements. The upper edge of the specimen is constrained such that no displacement occurs in the vertical direction. Rigid body motion is additionally restricted at the upper-left corner while still allowing horizontal contraction or expansion of the specimen. The vertical displacement of all nodes along the right edge is coupled to a single virtual node, whereas their horizontal displacements are unconstrained. The loading is applied to this virtual node using an arc-length procedure and the MCR effect is disabled.
		
		\paragraph{Discussion}
			The structural response predicted by the model is in good agreement with the experimental data (Fig.~\ref{fig:calibration1}). It lies within the scatter of the three experimental curves over the calibrated displacement range, indicating that the identified parameters provide a representative compromise. In particular, the elastic loading, the peak load, and the onset of softening are captured well. The subsequent decrease in load-carrying capacity is also reproduced reasonably, which demonstrates that the formulation can describe the early damage-driven softening of the L-shaped specimen. At a vertical displacement of approximately $0.4\,\si{\milli\meter}$, compressive states develop near the left edge of the specimen. The simulation is therefore terminated after this point due to the deactivated MCR effect in this setup. The identified model parameters constitute an additional result and are summarized in Tab.~\ref{tab:prototype_model_parameters} for further reference. These parameter values were used as the basis for the investigation of the gradient enhancement in section~\ref{ssec:exp_grad}, which complements the plain-concrete assessment to both quasi-static and cyclic loads.
			
	\subsection{Low-alloy steel -- model calibration and assessment against cyclic experiments}
		Experimental data, elastic parameters, and stress--cycle (S--N) curves describing the high-cycle fatigue regime of low-alloy steel were adopted from~\cite{noauthor_military_1990,golahmar_phase_2023}. The numerical data generated by the model of~\cite{golahmar_phase_2023} were used for calibration at four distinct stress amplitudes, while the remaining amplitudes were used for validation of the calibrated model at different load amplitudes. It is noted that the experimental data and the numerical reference contain data points that involve large stress amplitudes, which are typically not associated with high-cycle fatigue. Consequently, and due to the neglected plastic effects, the proposed model loses its applicability in this particular cycle range. Still, all data points are considered in the calibration and validation process in line with~\cite{golahmar_phase_2023} for completeness.
		
\begin{figure}[ht]
				\subfigreset
				\hspace{0.2cm}
				\subbfigure[Stress amplitudes versus cycles to failure (S--N diagram): comparison between calibration data and model prediction. $A = 1000 \, \si{\milli\meter^2}$ is the cross-sectional area.]{%
					\psfrag{g2}[c][c]{\scalebox{0.5}{\makebox{$b_\mr{initial} = 0.99$}}}
					\psfrag{L}[c][c]{\scalebox{0.5}{\makebox{$500\,\si{\milli\meter}$}}}
					\psfrag{uf}{\raisebox{0.1cm}{\scalebox{0.6}{\makebox{$\frac{\hat{f}}{A}$}}}}
					\psfrag{s}[c][c]{\scalebox{0.8}{\makebox{stress amplitude $\hat{\sigma} = \hat{f}/A \, [\si{\mega\pascal}]$}}}
					\psfrag{n}[c][c]{\scalebox{0.8}{\makebox{fatigue life $N_\mr{F} \, [-]$}}}
					\begin{tikzpicture}
						\node[anchor=south west,inner sep=0] (main) at (0,0)
						{\psfrag{DA}[l][c]{\scalebox{0.7}{experimental data~\cite{noauthor_military_1990}}}
							\psfrag{DB}[l][c]{\scalebox{0.7}{numerical reference~\cite{golahmar_phase_2023}}}
							\psfrag{DC}[l][c]{\scalebox{0.7}{model calibration}}
							\psfrag{DD}[l][c]{\scalebox{0.7}{model validation}}
							\includegraphics[width=0.45\textwidth]{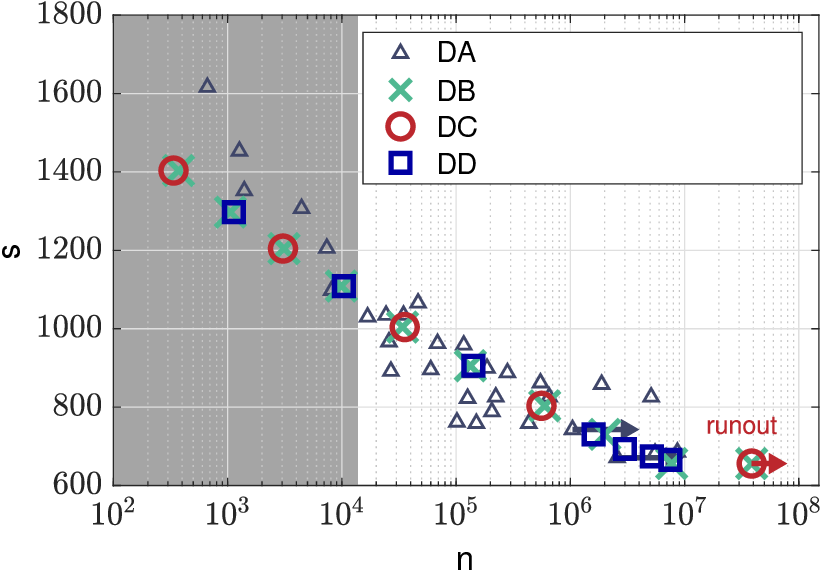}};
						\begin{scope}[x={(main.south east)},y={(main.north west)}]
							\node[anchor=north east] at (1.01,0.7) {\includegraphics[width=0.25\textwidth]{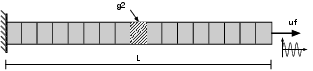}};
						\end{scope}
					\end{tikzpicture}
				}
			\hfill
				\subbfigure[Displacement amplitude versus cycle number for different applied stress amplitudes.]{%
					\psfrag{d}[c][c]{\scalebox{0.8}{\makebox{displacement amplitude $\hat{u} \, [\si{\milli\meter}]$}}}
					\psfrag{n}[c][c]{\scalebox{0.8}{\makebox{cycle number $N \, [-]$}}}
					\psfrag{g2}[c][c]{\scalebox{0.8}{\makebox{$b_0 = 0.99$}}}
					\psfrag{L}[c][c]{\scalebox{0.8}{\makebox{$\ell = 1\,\si{\milli\meter}$}}}
					\psfrag{A}[l][c]{\scalebox{0.8}{\makebox{$A = 1\,\si{\milli\meter^2}$}}}
					\psfrag{uf}{$F$}
					\raisebox{-1pt}{%
					\includegraphics[width=0.44\textwidth]{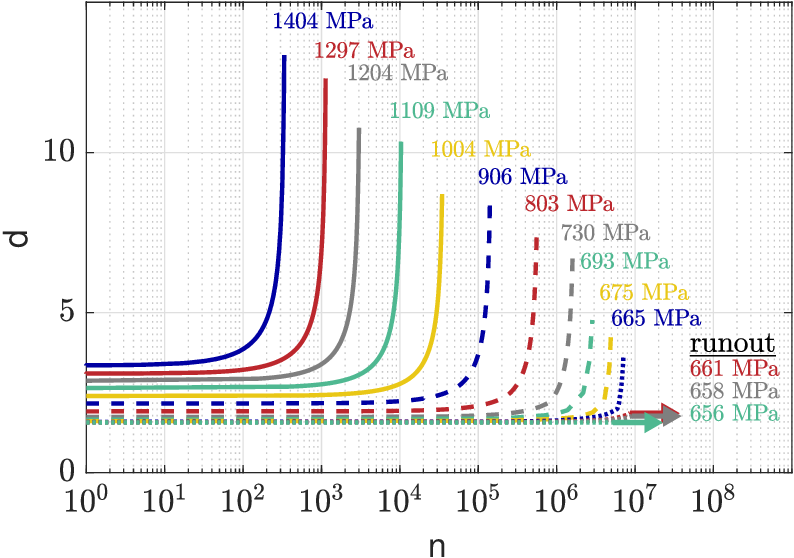}
				}
				}%
		\hfill
		\caption{Low-alloy steel -- model calibration and assessment against cyclic experiments: calibration of the fatigue model for 300M steel under cyclic loading. Predicted cycles to failure are compared with reference fatigue data from~\cite{golahmar_phase_2023}.}\label{fig:calibration2}
		\end{figure}

		\paragraph{Setup}
			A simplified one-dimensional simulation setup is considered, as illustrated in Fig.~\ref{fig:calibration2}. The bar specimen is discretized using $21$ linear finite elements and is clamped at one end while the opposite end is subjected to a cyclic stress amplitude. A cyclic uniaxial stress state is prescribed with a mean stress of $\sigma_\mr{mean} = 0 \,\si{\mega\pascal}$, corresponding to fully reversed loading (stress ratio $R=-1$). The MCR effect is also disabled during the calibration. However, as shown in Fig.~\ref{fig:d1rodresults:cyclic:academic:mcr}~(c) this simply implies a rescaling of the damage rate by a factor of $2$. An imperfection is introduced in the middle element by assigning an initial integrity value of $b_0 = 0.99$ to trigger localization. Consequently, this element exhibits the lowest integrity throughout the spatial domain during the simulation. The sensitivity with respect to this imperfection was checked by repeating the simulations of Fig.~\ref{fig:calibration2} with $b_0=0.999$. At stress amplitudes of $730\,\si{\mega\pascal}$ and $1004\,\si{\mega\pascal}$, the predicted fatigue lives increased by approximately $1.0\,\%$ and $0.5\,\%$, respectively. Thus, for the investigated cases, the predicted lifetime is only weakly affected by this variation, which does not change the findings of this study.
			
			A failure threshold of $b_\text{t} = 0.5$ is adopted for the present investigation. When the integrity of the middle element falls below this value, the specimen is considered to have failed. This corresponds to $25\,\% (= b_\mr{t}^2)$ of the initial stiffness remaining, owing to the quadratic appearance of the integrity tensor in Eq.~\eqref{eq:helmholtz}. Other threshold values can of course be chosen, for example, if required for other materials or by practical safety regulations. Yet, they would again lead to qualitatively similar results.
			
			To accelerate the repeated model evaluation during the calibration process, the internal variables are extrapolated over multiple cycles using a third-degree polynomial approximation whenever the estimated extrapolation error remains within a prescribed tolerance of $0.1\,\%$.
		
		\paragraph{Discussion}
			The resulting S--N curve of the calibrated model (Fig.~\ref{fig:d1rodresults}~(a)) is in good agreement with the numerical reference~\cite{golahmar_phase_2023} as well as the experimental data~\cite{noauthor_military_1990}. This match applies to the prescribed stress amplitudes of both the calibration and the validation data sets.
		As expected, the fatigue life decreases with increasing stress amplitude and follows an approximately linear trend in the logarithmic representation. The calibration results are summarized in Tab.~\ref{tab:prototype_model_parameters}. It should be noted that experimental high-cycle fatigue data are generally obtained using load-controlled servohydraulic or resonance-based testing machines, where the applied stress amplitude is prescribed. Under these conditions, simulations terminate as soon as strain softening occurs. Therefore, calibration requires the failure threshold $b_\mr{t}$ to be reached before softening begins, resulting in a relatively high damage exponent of $r=2.5$ due to its strong influence on softening through the integrity tensor. For example, a stress amplitude of $\sigma_\text{amp}=1404 \, \si{\mega\pascal}$ results in failure after approximately $372$ cycles ($336$ cycles predicted by the model and within a factor of $f=1.2$ of the reference value), whereas reducing the amplitude to $\sigma_\text{amp} = 730 \, \si{\mega\pascal}$ increases the fatigue life to more than $2\times10^6$ cycles (within a factor of $1.3$ from the reference value). The parameter $\eta_\text{i}$ was calibrated such that the model reproduces the experimentally observed runout behavior. As an important result, stress states below the pre-defined amplitude $\hat{\sigma}_\text{out} = 656 \, \si{\mega\pascal}$ fall inside the endurance surface and do not trigger further damage. Consequently, the calibrated model smoothly transitions from the finite-life regime of the S--N curve to the runout regime. The endurance surface is therefore not only a numerical threshold but also directly relates to the physical endurance limit through the calibration process.
		
		Turning to displacement amplitudes, the latter remain nearly constant initially (Fig.~\ref{fig:d1rodresults}~(b)). As damage accumulates, the material stiffness gradually decreases and increasingly larger displacements are required to sustain the loading. Consequently, the displacement amplitude increases rapidly once significant degradation develops, until the termination criterion is reached and the specimen is considered to have terminally failed. Two additional simulations were conducted near the transition to the runout regime at stress amplitudes of $\sigma_\text{amp} = 661 \, \si{\mega\pascal}$ and $\sigma_\text{amp}=658\,\si{\mega\pascal}$, respectively. As expected, the lower stress amplitudes produced smaller stable-phase displacement amplitudes, and only a slight increase in displacement amplitude was observed before the simulations reached the maximum cycle limit, leading to their classification as runout cases.
		
		These simulations demonstrate that the proposed framework and the endurance-surface concept can capture quasi-brittle degradation. In particular, damage evolution occurs over a large number of cycles, well within the regime commonly associated with high-cycle fatigue ($N > 10^5$), cf.~\cite{lemaitre_two_1999}.
		\begin{table}[ht]
			\def\arraystretch{1.2}
			\centering
			\begin{tabular}{| L{6.1cm} | C{1cm} | C{1.6cm} | C{1.6cm} | C{1.6cm} | C{1.0cm} |}
				\hline
				\multirow{3}{*}{Parameter name} & 
				\multirow{3}{*}{\!Symbol} & 
				\multicolumn{3}{c|}{Value} & 
				\multirow{3}{*}{\!Unit} \\ \cline{3-5}
				
				& &
				\multirow{2}{*}{\!Concrete} & 
				\multicolumn{2}{c|}{Steel} & 
				\\ \cline{4-5}
				
				& & &
				Isotropic & \!Anisotropic &
				\\
				
				\Xhline{1.5pt}
				First \text{Lam\'{e}} parameter & $\lambda$ & $6161$ & \multicolumn{2}{c|}{$121154$} & \si{\mega\pascal} \\ \hline
				Second \text{Lam\'{e}} parameter & $\mu$ & $10953$ & \multicolumn{2}{c|}{$80769$} & \si{\mega\pascal} \\ \hline
				Isotropic damage modulus & $\eta_\text{i}$ & $1.05 \times 10^7$ & $0.469$ & $0.328$ & \si{\mega\pascal^{-2}} \\ \hline
				Anisotropic damage modulus & $\eta_\text{a}$ & $0$ & $0$ & $0.141$ & \si{\mega\pascal^{-2}} \\ \hline
				Damage exponent & $r$ & $0.6$ & \multicolumn{2}{c|}{$2.51$} & -- \\ \hline 
				Damage scaling (exponential) & $K_1$ & -- & \multicolumn{2}{c|}{$10^{-10.88}$} & \si{\mega\pascal} \\ \hline 
				Damage scaling exponent (exponential) & $p_1$ & -- & \multicolumn{2}{c|}{$0.61$} & -- \\ \hline 
				Damage scaling (power law) & $K_2$ & $10^{-1.88}$ & \multicolumn{2}{c|}{$10^{-6.86}$} & \si{\mega\pascal} \\ \hline 
				Damage scaling exponent (power law) & $p_2$ & $6.195$ & \multicolumn{2}{c|}{$5.88$} & -- \\ \hline %
				Penalty parameter & $c_\text{b}$ & $5000$ & \multicolumn{2}{c|}{$15.59$} & \si{\mega\pascal} \\ \hline 
				Length scale parameter & $\ell_\text{b}$ & $0.25$ & \multicolumn{2}{c|}{\parbox[c]{4cm}{\centering\vspace{0.5mm}$150$ (cyclic 1D)\\ $1$ (roundbar)\vspace{0.5mm}}} & \si{\milli\meter} \\ \hline 
				Penalty exponent & $q$ & $0$ & \multicolumn{2}{c|}{$-1$}& --\\ 
				\hline
			\end{tabular}
			\caption{Prototype damage model: Model parameters identified for plain concrete (monotonic loading) and for 300M steel (cyclic loading, fatigue).}\label{tab:prototype_model_parameters}
		\end{table}

		\subsection{Low-alloy steel -- cylindrical roundbar specimen under axial--torsional loading}\label{ssec:2d_300m_cyclic}
\begin{figure}[ht]
				\centering
				\subfigreset
				\subbfigure[Machine setup~\cite{langenfeld_low_2023}, geometry of the axisymmetric specimen and applied loading configuration. Lengths are given in $\si{\milli\meter}$.]{%
				\psfrag{F}[l][c]{\scalebox{0.8}{\makebox{$\hat{F}$}}}
				\psfrag{u}[l][c]{\scalebox{0.8}{\makebox{$\hat{\gamma}$}}}
				\psfrag{l1}[c][c]{\scalebox{0.6}{\makebox{2.5}}}
				\psfrag{l2}[c][c]{\scalebox{0.6}{\makebox{5}}}
				\psfrag{R}[c][c]{\scalebox{0.6}{\makebox{R200}}}
				\includegraphics[width=0.4\textwidth]{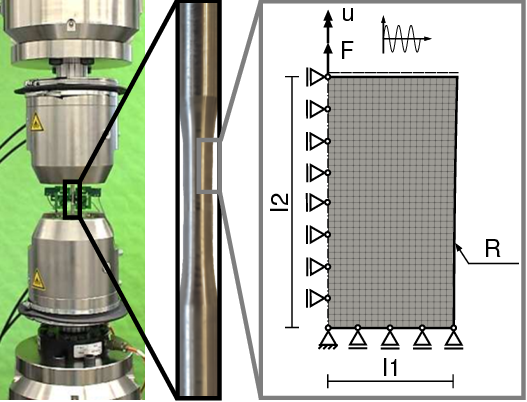}
				}
				\hspace{0.5cm}
				\subbfigure[Axial displacement and torque response versus number of cycles for the isotropic and anisotropic variant.]{%
					\psfrag{n}[c][c]{\scalebox{0.8}{\makebox{cycle number $N \, [-]$}}}
					\psfrag{trq}[c][c]{\scalebox{0.8}{\makebox{\textcolor{CS1}{torque $\hat{T} \, [\si{\newton\meter}]$}}}}
					\psfrag{dsp}[c][c]{\scalebox{0.8}{\makebox{\textcolor{CS2}{axial displacement $\hat{u} \, [\si{\micro\meter}]$}}}}
					\includegraphics[width=0.45\textwidth]{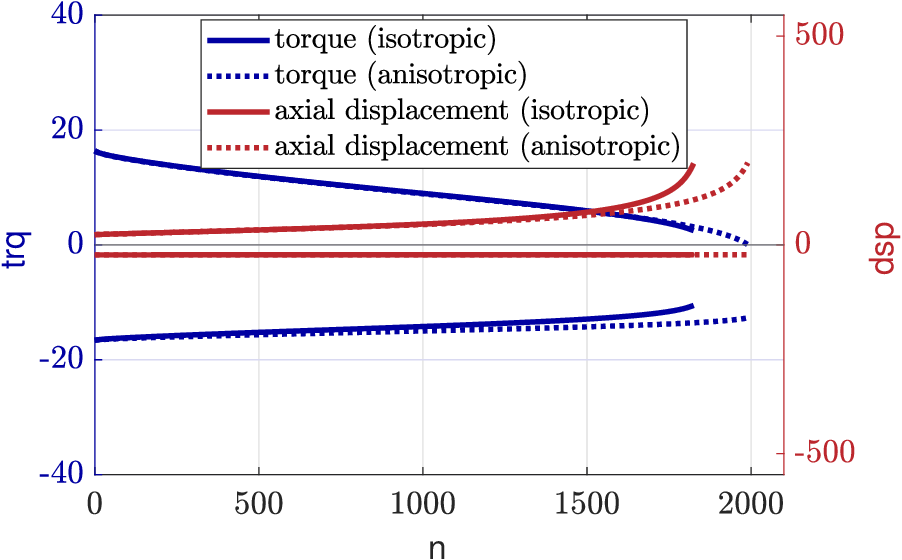}
				}

			\subbfigure[Integrity field of the isotropic (left, red) model after failure and differences of the anisotropic (right, blue) components to the isotropic integrity.]{%
			{\hspace{3ex}
				\psfrag{u}[r][l]{\scalebox{0.65}{\makebox{$0.56$}}}
				\psfrag{l}[r][l]{\scalebox{0.65}{\makebox{$0.54$}}}
				\psfrag{t}[c][c]{}
				\includegraphics[width=0.025\textwidth]{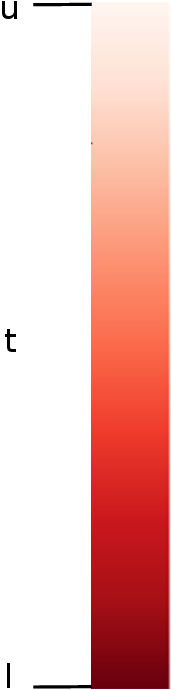}}
			\begin{tikzpicture}
				\node[anchor=south west, inner sep=0] (image) at (0,0)
				{%
					\includegraphics[width=0.132\textwidth]{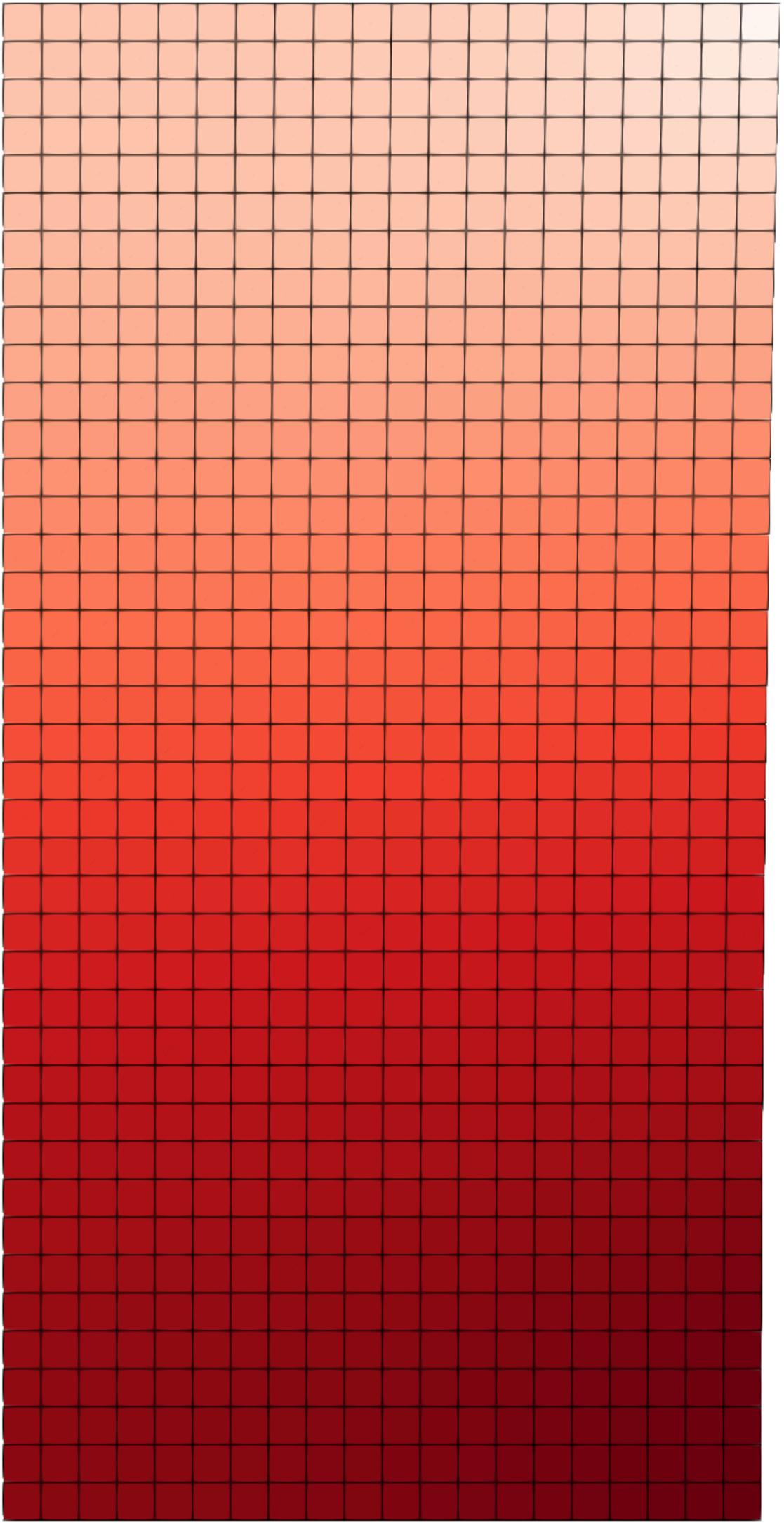}};
				\begin{scope}[x={(image.south east)}, y={(image.north west)}]
					\node[anchor=south west] at (-0.08, 0.87) {\scalebox{1}{\colorbox{white}{$b_\mr{iso}$}}};
					\node[anchor=south west] at (0, 0) {
						\psfrag{x}[c][c]{\textcolor{white}{$r$}}
						\psfrag{z}[c][c]{\textcolor{white}{$z$}}
						\psfrag{y}[c][c]{\textcolor{white}{$\theta$}}
						\includegraphics[width=0.08\textwidth]{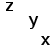}};
				\end{scope}
			\end{tikzpicture}
			\hspace{12ex}
			\begin{tikzpicture}
				\node[anchor=south west, inner sep=0] (image) at (0,0)
				{%
					\includegraphics[width=0.132\textwidth]{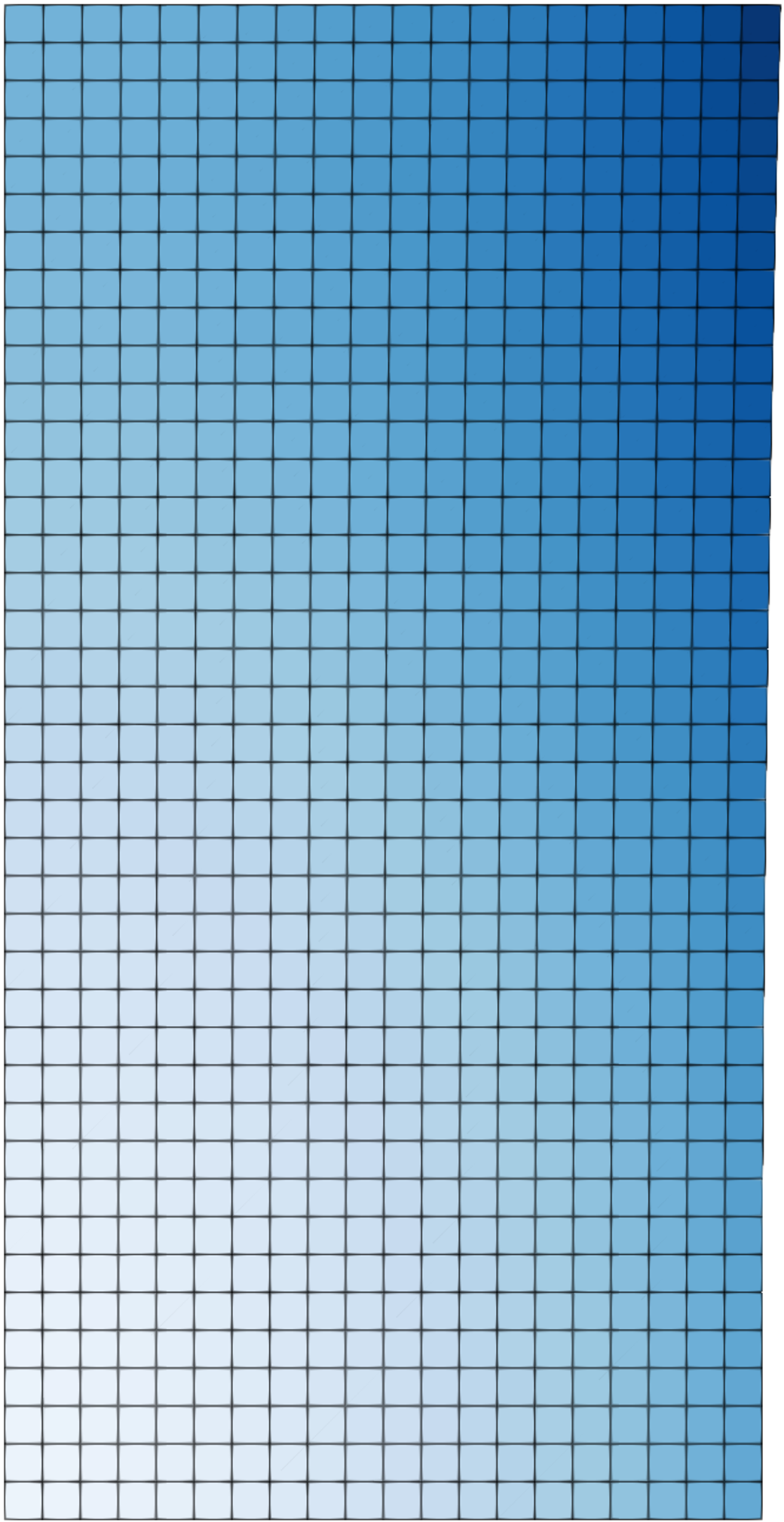}};
				\begin{scope}[x={(image.south east)}, y={(image.north west)}]
					\node[anchor=south west] at (-0.08, 0.87) {\scalebox{1}{\colorbox{white}{$b_\mr{z}-b_\mr{iso}$}}};
				\end{scope}
			\end{tikzpicture}
			{\psfrag{u}[l][r]{\scalebox{0.65}{$+1.25\,\%$}}
				\psfrag{l}[l][r]{\scalebox{0.65}{$-1.25\,\%$}}
				\psfrag{t}[l][r]{\scalebox{0.65}{$0.1\,\%$}}
				\includegraphics[width=0.025\textwidth]{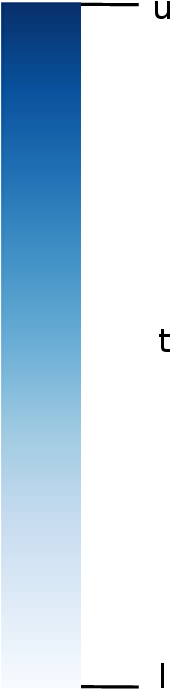}\hspace{5ex}}
			\begin{tikzpicture}
				\node[anchor=south west, inner sep=0] (image) at (0,0)
				{%
					\includegraphics[width=0.132\textwidth]{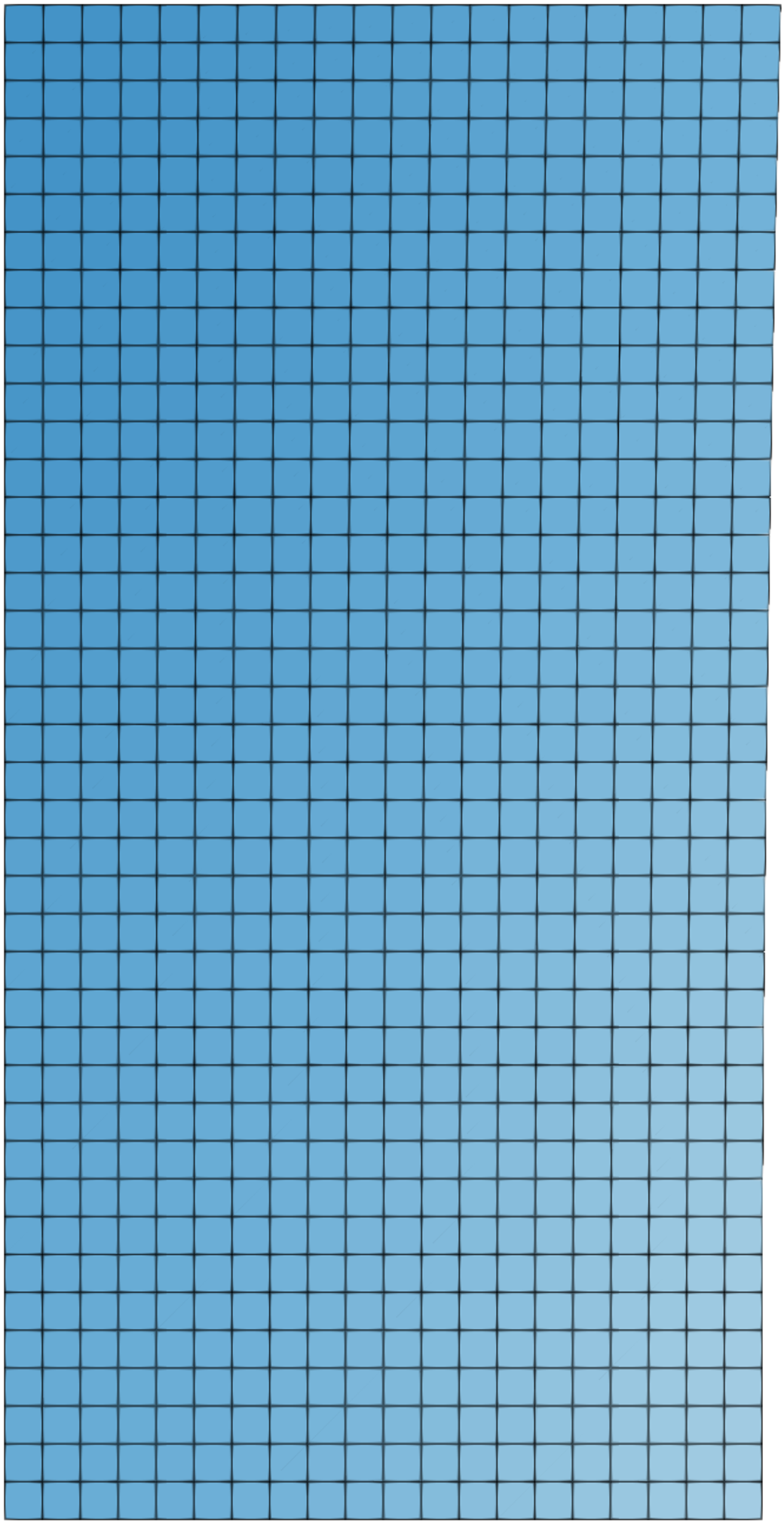}};
				\begin{scope}[x={(image.south east)}, y={(image.north west)}]
					\node[anchor=south west] at (-0.08, 0.87) {\scalebox{1}{\colorbox{white}{$b_\mr{r}-b_\mr{iso}$}}};
				\end{scope}
			\end{tikzpicture}
			{\psfrag{u}[l][r]{\scalebox{0.65}{$+1.25\,\%$}}
				\psfrag{l}[l][r]{\scalebox{0.65}{$-1.25\,\%$}}
				\psfrag{t}[l][r]{\scalebox{0.65}{$18.6\,\%$}}
				\includegraphics[width=0.025\textwidth]{blues_scale_t}\hspace{5ex}}
			\begin{tikzpicture}
				\node[anchor=south west, inner sep=0] (image) at (0,0)
				{%
					\includegraphics[width=0.132\textwidth]{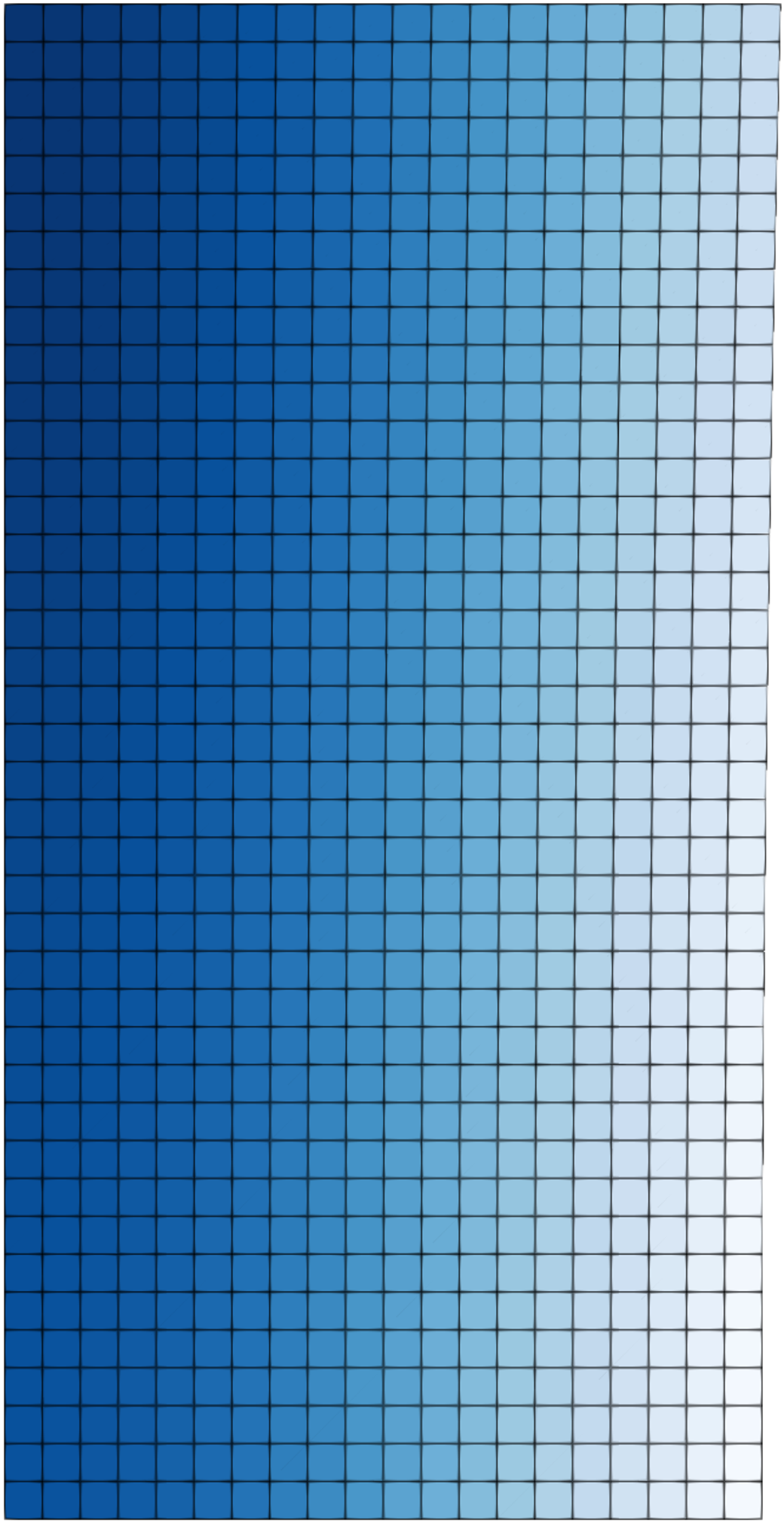}};
				\begin{scope}[x={(image.south east)}, y={(image.north west)}]
					\node[anchor=south west] at (-0.08, 0.87) {\scalebox{1}{\colorbox{white}{$b_\theta-b_\mr{iso}$}}};
				\end{scope}
			\end{tikzpicture}
			{\psfrag{u}[l][r]{\scalebox{0.65}{$+1.25\,\%$}}
				\psfrag{l}[l][r]{\scalebox{0.65}{$-1.25\,\%$}}
				\psfrag{t}[l][r]{\scalebox{0.65}{$17.7\,\%$}}
				\includegraphics[width=0.025\textwidth]{blues_scale_t}\hspace{6ex}}
			}
			\caption{Low-alloy steel -- cylindrical roundbar specimen under axial--torsional loading: geometry, boundary conditions, and structural response under cyclic axial--torsional loading for the isotropic and an anisotropic variant.}\label{fig:rb2d_setup_fu}
		\end{figure}

		This example focuses on the comparison between isotropic and anisotropic material degradation for a cylindrical roundbar specimen under cyclic axial--torsional loading (Fig.~\ref{fig:rb2d_setup_fu}).
		
		\paragraph{Setup}
			A two-dimensional axisymmetric slice of the specimen is discretized using 800 quadrilateral elements (Fig.~\ref{fig:rb2d_setup_fu}~(a)). The specimen has a slight taper toward its center to define a unique region for promoted damage accumulation. Axial and horizontal symmetry conditions are exploited in the numerical model. Additional geometric constraints ensure that the upper and lower edges remain planar and horizontal during loading.
			
			Through these constraints, the upper edge is subjected to cyclic axial loading with a force amplitude of $\hat{F} = 20.25 \, \si{\kilo\newton}$ combined with an in-phase torsional rotation of amplitude $\hat{\gamma} = 0.016125 \, \si{\radian} \approx 0.92 \, \si{\degree}$. In addition to the isotropic parameter set, an artificial anisotropic variant is considered by redistributing the damage modulus from $\eta_\mr{i}$ to $\hat\eta_\mr{i} = 0.7\,\eta_\mr{i}$ and $\hat\eta_\mr{a} = 0.3\,\eta_\mr{i}$ as one possible parameter perturbation for demonstration. In both cases, the MCR effect and finite strain kinematics~\cite{GreenNaghdi1965} are considered. To find a compromise between computational effort and cyclic loading, the applied loads are chosen such that the experiments result in $1800$ to $2000$ cycles to failure. The length-scale parameter has been modified to $\ell_\mr{b}=1\,\si{\milli\meter}$ in order to account for the size of the specimen.
			
		\paragraph{Discussion}
			The numerical results reveal a clear tension--compression asymmetry in both the torque and displacement responses (Fig.~\ref{fig:rb2d_setup_fu}~(b)).
		The axial displacement remains nearly constant during the compression phases for both the isotropic and anisotropic variants. In contrast, the displacement amplitude increases progressively during the tensile phases. The torque response shows a more subtle distinction between tension and compression. Its reduction is more pronounced when torsion is superposed with tensile axial loading than with compressive axial loading.
		
		In the final hundreds of cycles, the overall degradation accelerates, leading to a rapid increase in displacement before failure. The model including anisotropic degradation exhibits a delayed overall degradation compared with the isotropic variant. The anisotropic formulation terminated at $1990$ cycles, whereas the isotropic simulation terminated at $1825$. This corresponds to a $9\,\%$ increase in lifetime by allowing for an anisotropic integrity distribution.
		
		This observation is also reflected in the integrity fields of both model variants (Fig.~\ref{fig:rb2d_setup_fu}~(c)). The integrity of the isotropic model generally shows a relatively homogeneous distribution over the specimen owing to the absence of strain localization. However, lower integrity concentrates toward the tapered middle of the specimen, with a nearly constant distribution over the radial coordinate.
		
		The differences between the anisotropic and the isotropic model variants reveal more intricate details. First, the tensor component in the $z$-direction lies within the value range of the isotropic integrity. It is, however, more heterogeneously distributed along the radial axis. The circumferential ($\theta$-)direction also shows a much more heterogeneous distribution along the radial axis. However, the more significant observation is the large deviation in the radial and circumferential integrity tensor components relative to the isotropic case. These components are less damaged (they retain about $19\,\%$ more integrity), which is one reason for the prolonged predicted lifetime in the anisotropic case. It should be noted that the off-diagonal entries of the anisotropic integrity tensor are sufficiently small such that the eigenvalues closely follow the main-diagonal entries and the eigenvectors are closely aligned with the provided coordinate vectors.
		
		This numerical experiment eventually demonstrates the practical relevance of anisotropic degradation in the near-failure regime. While only minor to moderate differences between the isotropic and anisotropic models are observed during the early and intermediate stages of loading, pronounced deviations emerge shortly before failure, particularly in the local integrity fields as well as in the global torque and displacement responses.

\section{Conclusion}\label{sec:conclusion}
	This work develops a physically sound model for the prediction of quasi-brittle damage based on the concept of an endurance surface. The benefit of this concept is the straightforward distinction between elastic and inelastic states, making it particularly suitable from monotonic to cyclic loads. Damage evolves driven by the elastic energy of the material and depends on the distance from an endurance surface. The technical capabilities include: a micromorphic enhancement for regularization, the MCR effect for a realistic tension--compression asymmetry, and anisotropic damage evolution for strongly directional load conditions during the service life.

	A key outcome of the study is that the proposed endurance-surface formulation captures both monotonic and cyclic degradation as demonstrated for plain concrete and low-alloy steel. Successful calibration to both materials is enabled by the proposed exponential-plus-power-law type scaling function. For plain concrete, the elastic loading, the peak load, and the softening of a monotonically loaded L-shaped specimen captured experimental reference data well. For low-alloy steel, the S--N curve of a bar specimen was captured over a large range of stress amplitudes and up to $10^7$ cycles, successfully coupled with a polynomial cycle-extrapolation. Furthermore, the behavior of the model under complex loading conditions was investigated using an axisymmetric cylindrical specimen subjected to combined axial--torsional loading at finite strains and by considering the MCR effect. The cyclic experiment compared isotropic and anisotropic degradation, which showed that introducing anisotropy can influence the fatigue life, most notably in the near-failure regime.
	
	The proposed formulation of the endurance surface was physically motivated through the energy-release rate as driving forces and led to a robust modeling framework. A potential-based approach for the evolution equation allowed for anisotropic degradation in simulations within the thermodynamically consistent GSM framework. The incorporation of a gradient enhancement indicated self-stabilizing behavior toward coupling the integrity and auxiliary fields, and was shown to eliminate mesh dependence and provide stable, physically meaningful damage fields. Numerical examples confirmed the robustness of the approach, demonstrating convergence of the macroscopic responses for increasing mesh refinements.

	Overall, the proposed framework provides a thermodynamically consistent and computationally robust approach for the numerical simulation of quasi-brittle material degradation. It is applicable to the estimation of the ultimate load under monotonic loading as well as the service-life under cyclic loading. The practical relevance of the proposed framework is demonstrated by addressing three representative benchmark settings: the L-shaped concrete specimen as a well-established test for monotonic quasi-brittle failure, S--N curves as the classical basis for fatigue-life assessment, and axial--torsional loading as a laboratory-relevant scenario for imposing complex multiaxial load paths and inducing anisotropic degradation in a numerical experiment.

	While the present study focuses on demonstrating the theoretical framework and numerical capabilities of the endurance-surface formulation, further validation against dedicated experimental multiaxial fatigue data will be addressed in future work. In particular, the interaction between anisotropic fatigue degradation and nonproportional loading paths will be investigated experimentally, adding to the present purely numerical investigation. Additional developments will address the coupling with constitutive models suitable for ductile damage in the low-cycle fatigue regime as well as computational strategies for efficiently simulating very large numbers of load cycles.

\newpage
\appendix
\section{Finite element discretization}\label{ssec:finite_element_implementation}
	This section outlines the finite element implementation following~\cite{langenfeld_micromorphic_2020}. The displacement field $\B{u}$ and the micromorphic field $\B{\varphi}$, together with their variations, are discretized using $C^0$ finite elements following the Bubnov--Galerkin-Ansatz. The approximations read
	\begin{align}
		\B{u} &= \sum_{A=1}^{n_\text{en}} \B{u}^A \, N^A\,\text{,} \quad
		\B{\varphi} = \sum_{A=1}^{n_\text{en}} \B{\varphi}^A \, N^A \, \text{,} \quad
		\delta\B{u} = \sum_{A=1}^{n_\text{en}} \delta \B{u}^A \, N^A\,\text{,} \quad
		\delta\B{\varphi} = \sum_{A=1}^{n_\text{en}} \delta \B{\varphi}^A \, N^A
	\end{align}
	where $\B{u}^A$ and $\B{\varphi}^A$ denote the discrete nodal values associated with node $A$. $N^A$ is the corresponding shape function. The gradients of the fields follow as
	\begin{align}
		\nabla \B{u} &= \sum_{A=1}^{n_\text{en}} \B{u}^A \otimes \partDer{N^A}{\B{X}}\,\text{,} \quad
		\nabla \B{\varphi} = \sum_{A=1}^{n_\text{en}} \B{\varphi}^A \otimes 	\partDer{N^A}{\B{X}}\,\text{,}\\
		\nabla \delta\B{u} &= \sum_{A=1}^{n_\text{en}} \delta\B{u}^A \otimes 	\partDer{N^A}{\B{X}}\,\text{,} \quad
		\nabla \delta\B{\varphi} = \sum_{A=1}^{n_\text{en}} \delta\B{\varphi}^A \otimes 	\partDer{N^A}{\B{X}}\,\text{.}
	\end{align}
	The residual contributions associated with element $e$ (neglecting body forces) read
	\begin{align}
		\delta \B{u}^A \cdot \B{R}^{e,A}_{\B{u}} &= \int_{V^e} \B{\sigma} : \delta \B{\ve} \, \mr{d}V = \int_{V^e} 	\B{\sigma} : \nabla \delta \B{u}^A \, \mr{d}V = \delta \B{u}^A \cdot \int_{V^e} \B{\sigma} \cdot \partDer{N^A}{\B{X}} \, \mr{d}V\, \text{,}\label{eq:FEM_residual_u}
		\\
		\delta \B{\varphi}^A \cdot \B{R}^{e,A}_{\B{\varphi}} &= \int_{V^e} \B{\omega} : \delta \B{\varphi} + \B{\Omega} \cccon 	\nabla\delta\B{\varphi} \, \mr{d}V = \delta \B{\varphi}^A \cdot \int_{V^e} \left( \B{\omega}\,N^A + \B{\Omega} \cdot \partDer{N^A}{\B{X}} \right) \, \mr{d}V\,\text{.}\label{eq:FEM_residual_phi}
	\end{align}

	The resulting nonlinear system is solved using a Newton--Raphson scheme, yielding the displacement and micromorphic fields. Linearization of the residuals~\eqref{eq:FEM_residual_u} and~\eqref{eq:FEM_residual_phi} yields
	\begin{align}
		\B{K}^{e,AB}_{\B{u}\B{u}} &= \totDer{\B{R}^{e,A}_{\B{u}}}{\B{u}^B} = \int_{V_e} \partDer{N^A}{\B{X}} \cdot \totDer{\B{\sigma}}{\B{\ve}} \cdot \partDer{N^B}{\B{X}} \, \mr{d}V \, \text{,}
		\\
		\B{K}^{e,AB}_{\B{u}\B{\varphi}} &= \totDer{\B{R}^{e,A}_{\B{u}}}{\B{\varphi}^B} = \int_{V_e} \partDer{N^A}{\B{X}} \cdot \totDer{\B{\sigma}}{\B{\varphi}} \, N^B \, \mr{d}V \, \text{,} \quad
		\B{K}^{e,AB}_{\B{\varphi}\B{u}} = \totDer{\B{R}^{e,A}_{\B{\varphi}}}{\B{u}^B} = \int_{V_e} N^A \, \totDer{\B{\omega}}{\B{\ve}} \cdot \partDer{N^B}{\B{X}} \, \mr{d}V \, \text{,}
		\\
		\B{K}^{e,AB}_{\varphi_{ij}\varphi_{kl}} &= \totDer{\B{R}^{e,A}_{\varphi_{ij}}}{\varphi^B_{kl}} = \int_{V_e} N^A \, \left[\totDer{\B{\omega}}{\B{\varphi}}\right]_{ijkl} \, N^B + \left[\partDer{N^A}{\B{X}}\right]_m \, \left[\totDer{\B{\Omega}}{\nabla \B{\varphi}}\right]_{ijmkln} \, \left[\partDer{N^B}{\B{X}}\right]_n \, \mr{d}V \, \text{.}
	\end{align}
	\begin{align}
		\mr{d}\B{\sigma} &= \left[ \partDer{\B{\sigma}}{\B{\ve}} + \partDer{\B{\sigma}}{\B{b}} : \partDer{\B{b}}{\B{\ve}} \right] : \mr{d}\B{\ve} + \partDer{\B{\sigma}}{\B{b}} : \partDer{\B{b}}{\B{\varphi}} : \mr{d}\B{\varphi} \, \text{,}
		\\
		\mr{d}\B{\omega} &= \left[ \partDer{\B{\omega}}{\B{b}} : \partDer{\B{b}}{\B{\ve}} \right] : \mr{d}\B{\ve} + \left[ \partDer{\B{\omega}}{\B{\varphi}} + \partDer{ \B{\omega} }{ \B{b} } : \partDer{\B{b}}{\B{\varphi}} \right] : \mr{d}\B{\varphi}\, \text{,} \quad
		\mr{d}\B{\Omega} = \partDer{\B{\Omega}}{\nabla \B{\varphi}} : \mr{d} \nabla \B{\varphi}\, \text{,}
	\end{align}
	leading to the consistent algorithmic tangent. The state variable $\B{b}$ is computed locally at each integration point using a return-mapping algorithm solved by a Newton--Raphson iteration. The residual corresponding to the implicit evolution equation reads
	\begin{align}
		\B{r}\of{\B{\ve}, \B{\varphi}, \B{b}} &= \B{b} - \B{b}_n - \mathcal{H}\of{ f } \, \mac{ f - f_n} \, \gamma\of{ f } \, \tfrac{\partial \Gamma\of{\B{\beta}}}{\partial \B{\beta}},\\
		\text{with} \quad f &= f\of{\B{\beta}\of{\B{\ve},\, \B{b}}, \, \B{b}} \quad \text{and} \quad f_n = f\of{\B{\beta}\of{\B{\ve}_n,\, \B{b}_n}, \, \B{b}_n}
	\end{align}
	where the index $\bullet_n$ denotes the last converged solution, while variables without index refer to the current solution. The update of $\B{b}$ is obtained by using Newton's method,
	\begin{align}
		\B{b}_{k+1} \leftarrow \B{b}_k - \left[ \left. \B{J} \right\vert_{\B{b}=\B{b}_k} \right]^{-1} \circ \left. \B{r} \right\vert_{\B{b}=\B{b}_k}
	\end{align}
	where $\circ$ denotes the appropriate tensor contraction. Index $k$ denotes the current iteration and $\B{J} \coloneqq \tfrac{\partial\B{r}}{\partial\B{b}}$ is the Jacobian of the residual vector. Linearization of the residual at the converged solution together with the consistency condition $\mr{d}\B{r} = \B{0}$ yields
	\begin{align}
		\mr{d}\B{r} &= \partDer{\B{r}}{\left[\B{\ve}, \B{\varphi} \right]} \circ \mr{d} \! \begin{bmatrix}
			\B{\ve}\\
			\B{\varphi}
		\end{bmatrix} + \B{J} \circ \mr{d}\B{b} = \B{0} \quad
		\Leftrightarrow \quad \mr{d}\B{b} = - \B{J}^{-1} \circ \partDer{\B{r}}{\left[ \B{\ve}, \B{\varphi} \right]} \circ \mr{d} \! \begin{bmatrix}
			\B{\ve}\\
			\B{\varphi}
		\end{bmatrix}
	\end{align}
	from which the sensitivities follow as
	\begin{align}
		\totDer{\B{b}}{\B{\ve}} &= - \B{J}^{-1} \circ \partDer{\B{r}}{\B{\ve}} \, \text{,} \quad 
		\totDer{\B{b}}{\B{\varphi}} = - \B{J}^{-1} \circ \partDer{\B{r}}{\B{\varphi}} \, \text{.}
	\end{align}

	\paragraph{Finite-strain extension}
	The numerical experiment in~\cref{ssec:2d_300m_cyclic} is based on a finite-strain formulation. Based on the kinematics described in section~\ref{ssec:framework}, the deformation gradient $\B{F}$ and the logarithmic strain measure $\B{H}$ are introduced as
	\begin{align}
		\B{F} &= \partDer{\left[\B{X}+\B{u}\of{\B{X}}\right]}{\B{X}} = \nabla_{\!\B{X}} \B{u} + \B{I}, \quad \B{H} = \frac{1}{2}\,\text{ln}\of{\B{F}^\text{t} \cdot \B{F}}
	\end{align}
	where the tensor logarithm is evaluated using the spectral decomposition. The finite element formulation is modified accordingly to account for the resulting nonlinear strain measure.
	
	The computation of the displacement gradient is further adapted to account for axisymmetry using cylindrical coordinates ($R, Z, \Theta$) in the reference configuration, resulting in the gradient operator $\nabla_{\!\left[R,Z,\Theta\right]} \B{u}$. The displacements in the current (spatial) configuration are computed via
	\begin{align}
		\begin{bmatrix}
			r\of{R,Z}\\
			z\of{R,Z}\\
			\theta\of{R,Z,\Theta}
		\end{bmatrix} = \begin{bmatrix}
			R+u_\mathrm{R}\of{R,Z}\\
			Z+u_\mathrm{z}\of{R,Z}\\
			\Theta+u_\Theta\of{R,Z}
		\end{bmatrix}, \quad \text{with} \,\, \nabla_{\!\left[R,Z,\Theta\right]} = \begin{bmatrix} \partDer{}{R}, \quad \partDer{}{Z}, \quad \frac{1}{R}\,\partDer{}{\Theta}
		\end{bmatrix}\,\text{.}
	\end{align}
	The nodal values of the micromorphic field are interpolated to the Gauss points using the same shape functions as described in~\ref{ssec:finite_element_implementation}. These values are coupled to the local integrity tensor. The gradient of the micromorphic field is computed consistently in the cylindrical coordinate system introduced above. The formulation corresponds to the standard axisymmetric finite element setting. Within the framework described in~\cref{ssec:framework}, the linear strain tensor $\B{\ve}$ is replaced by the logarithmic strain tensor $\B{H}$. Apart from these modifications, the formulation remains unchanged.

\section{Model details}

	\subsection{Strain energy equivalence framework for anisotropic ductile damage}\label{ssec:framework:o}
		\renewcommand{\ieps}{\B{b} : \B{\ve}\el_+ + \B{I}:\B{\ve}\el_-}
		\renewcommand{\iieps}{\B{b} : \left[ \B{\ve}\el_+ \cdot \B{b} \cdot \B{\varepsilon}\el_+\right] + \B{\ve}\el_-:\B{\ve}\el_- }
		\newcommand{\ikk}{\B{b} : \B{k}}
		\newcommand{\iia}{\B{b} : \left[ \B{a} \cdot \B{b} \cdot \B{a} \right]}
		\newcommand{\iiepsp}{\B{b} : \left[ \B{\ve}\pl \cdot \B{b} \cdot \B{\ve}\pl \right]}
		
		This work extends a material model based on the principle of strain energy equivalence~\cite{STEINMANN19981793}. The constitutive relations follow the formulations proposed in~\cite{Menzel02, Ekh03}. The Helmholtz energy $\psi$ is additively decomposed into an elastic part $\psi\el$ and a plastic part $\psi\pl$,
		\begin{align}
			\psi &= \psi\el(\B{\ve}, \, \B{\ve}\pl, \, \B{b}) + \psi\pl( \B{a}, \, \B{k}, \, \B{b}, \, \B{\ve}\pl) \, \text{,} \label{eq:helmholtz:o}
			\\
			\psi\el &= \dfrac{\lambda}{2} \, \left[\ieps\right]^2 + \mu \, \left[\iieps\right] \, \text{,}
			\\
			\psi\pl &= \dfrac{H_i}{2} \, \left[\ikk \right]^2 + \dfrac{H_\mr{a}}{2} \, \iia \, \text{.}
		\end{align}
		A small-strain setting is employed, which allows for the additive decomposition of the strain tensor $\B{\ve}$ into elastic and plastic parts, i.e., $\B{\ve} = \B{\ve}\el + \B{\ve}\pl$~\cite{GreenNaghdi1965}. Isotropic and kinematic hardening are captured through the state variables $\B{k}$ and $\B{a}$, respectively. For consistency with the principle of strain energy equivalence, isotropic hardening is described using a tensorial internal variable~\cite{Ekh03}. The corresponding parameters are $H_\mr{i}$ (isotropic hardening modulus) and $H_\mr{a}$ (nonlinear kinematic hardening modulus).
		
		The microcrack-closure-reopening (MCR) effect is incorporated to account for tension--compression asymmetry in the damage evolution. For this purpose, the elastic strain tensor is additively decomposed into tensile and compressive parts,
		\begin{align}\label{eq:mcr_decomp:o}
			\B{\ve}\el = \B{\ve}\el_{+} + \B{\ve}\el_{-} \, \text{,} \quad \B{\ve}\el_{+} &= \sum_{i=1}^{3} \mathcal{H}\of{\ve\el_{i}} \, \ve\el_{i} \, \B{N}_{i}^\ve \otimes \B{N}_{i}^\ve \, \text{,}
		\end{align}
		where $\ve\el_i$ denote the eigenvalues and $\B{N}_i^\ve$ the associated eigenvectors of the elastic strain tensor. $\mathcal{H}$ denotes the Heaviside function, which has been approximated following~\cite{Ekh03} by choosing numerical parameters $g_0 = 0$, $x_0 = 0$ and $x_R = 10^{-6}$. The thermodynamic driving forces follow from Helmholtz energy~\eqref{eq:helmholtz:o} as
		\begin{align}
			\B{\sigma}\phantom{\el} &=\phantom{-}\partDer{\psi\phantom{\el}}{\B{\ve}} = \lambda \, \left[\ieps\right] \, \left[ \B{b} : \partDer{\B{\ve}\el_+}{\B{\ve}\el} +
			\B{I} : \partDer{\B{\ve}\el_-}{\B{\ve}\el}\right] + 2\,\mu\, \left[ \left[ \B{b} \cdot \B{\ve}\el_+ \cdot \B{b} \right] : \partDer{\B{\ve}\el_+}{\B{\ve}\el} +
			\B{\ve}\el_- : \partDer{\B{\ve}\el_-}{\B{\ve}\el} \right] \, \text{,}
			\\
			\B{\alpha}\phantom{\el} &= -\partDer{\psi\phantom{\el}}{\B{a}} = - H_a\, \B{b} \cdot \B{a} \cdot \B{b}\, \text{,}
			\\
			\B{\kappa}\phantom{\el} &= -\partDer{\psi\phantom{\el}}{\B{k}} = - H_i \, \left[ \B{b} : \B{k} \right]\, \B{b}\, \text{,}
			\\
			\B{\beta}\el &= -\partDer{\psi\el}{\B{b}} = - \lambda \, \left[ \B{b} : \B{\ve}\el_+ + \B{I} : \B{\ve}\el_- \right] \B{\ve}\el_+ -
			2 \, \mu \, \B{\ve}\el_+ \cdot \B{b} \cdot \B{\ve}\el_+ \, \text{,}\label{eq:plasti_bet_el}
			\\
			\B{\beta}\pl &= -\partDer{\psi\pl}{\B{b}} = - H_a \, \B{a} \cdot \B{b} \cdot \B{a} - H_i \, \left[ \B{b} : \B{k} \right] \, \B{k} \, \text{,}
			\\
			\B{\beta}\phantom{\el} &= \B{\beta}\el + \B{\beta}\pl \, \text{,}
		\end{align}
		where $\B{\sigma}$ denotes the stress tensor, $\B{\alpha}$ the back stress tensor related to kinematic hardening, $\B{\kappa}$ the drag stress tensor related to isotropic hardening, and $\B{\beta}$ the energy-release-rate tensor.
		
		A plastic potential is introduced in line with the generalized standard materials framework~\cite{Halphen1975SurLM} in order to derive the evolution equations. The dissipation inequality is automatically satisfied if the potential is convex, nonnegative and contains the origin. The prototype plastic potential is defined as
		\begin{align}
			g &= \Phi(\B{\sigma}, \, \B{\alpha}, \, \B{\kappa, \, \B{b}}) + \Gamma_\alpha(\B{\alpha}, \, \B{b}) + \Gamma_\beta(\B{\beta}\el, \, \B{b}) \label{eq:plastpot:o}
		\end{align}
		and consists of the yield function $\Phi$ together with the nonassociative contributions $\Gamma_{\!\alpha}$ and $\Gamma_\beta$. The yield function follows~\cite{langenfeld_low_2023} and is defined as
		\begin{alignat}{3}
			&&&\Phi &&= \sqrt{\bar{\tau}} - \tau_y - \frac{1}{3} \, \B{b}^{-1} : \B{\kappa} - \Delta \tau \, \left[ 1 - \exp{-\frac{\left\vert \B{b}^{-1} : \B{\kappa} \right\vert}{\kappa_\mr{u}}} \right] \, \text{,} \label{eq:yield:o}
			\\
			& &&\bar{\tau} &&= \dfrac{3}{2} \, \B{b}^{-1} : \left[ \B{\tau} \cdot \B{b}^{-1} \cdot \B{\tau} \right] - \dfrac{1}{2} \, \left[\B{b}^{-1} : \B{\tau} \right]^2 \quad \text{with} \quad \B{\tau} = \B{\sigma} - \B{\alpha} \, \text{.}
		\end{alignat}
		Here, the von Mises--type equivalent stress $\sqrt{\bar{\tau}}$ depends on the relative stress tensor $\B{\tau}$. The yield function captures linear and exponential isotropic hardening through the direct incorporation of the drag stress tensor $\B{\kappa}$. The corresponding model parameters are $\tau_y$, $\Delta \tau$ and $\kappa_\mr{u}$. The nonassociative contributions of potential~\eqref{eq:plastpot:o} read
		\begin{align}
			\Gamma_\alpha &= \dfrac{B_\mr{a}}{2 \, H_\mr{a}} \, \B{b}^{-1} : \left[ \B{\alpha} \cdot \B{b}^{-1} \cdot \B{\alpha} \right] \, \text{,}
			\\
			\Gamma_\beta &= \dfrac{C_\mr{i}}{2} \, \left[\B{b}^m : \B{\beta}\el\right]^2 + \dfrac{C_\mr{a}}{2} \, \B{b}^m : \left[ \B{\beta}\el \cdot \B{b}^m \cdot \B{\beta}\el \right] \, \text{.} \label{eq:gamma_beta:o}
		\end{align}
		Nonlinear kinematic hardening of Armstrong--Frederick type is incorporated through parameter $B_\mr{a}$, cf.~\cite{frederick_mathematical_2007}. Damage evolution is governed by the nonassociative potential $\Gamma_\beta$, which is formulated in terms of the elastic part of the energy-release rate, $\B{\beta}\el = \B{\beta} - \B{\beta}\pl$, rather than $\B{\beta}$. This represents a deliberate modeling choice and leads to damage evolution being driven by the elastic energy. Isotropic damage evolution is controlled by the parameter $C_\mr{i}$, whereas the anisotropic damage evolution is governed by $C_\mr{a}$. The exponent $m$ provides additional flexibility for calibrating the model to experimental data~\cite{Ekh03}.
		
		The evolution equations follow in a straightforward manner as gradients of the plastic potential~\eqref{eq:plastpot:o} and read
		\begin{align}\label{eq:evolution_plasti:o}
			\dot{\B{\varepsilon}}\pl = \dot{\lambda} \, \partDer{g}{\B{\sigma}}, \quad \dot{\B{a}} = \dot{\lambda} \, \partDer{g}{\B{\alpha}}, \quad \dot{\B{k}} = \dot{\lambda} \, \partDer{g}{\B{\kappa}}, \quad \dot{\B{b}} = \dot{\lambda} \, \partDer{g}{\B{\beta}} = \dot{\lambda} \, \partDer{g}{\B{\beta}\el} \, \text{.}
		\end{align}
		This formulation is consistent with the generalized standard materials framework.
		
		Note that choosing $C_\mr{a} = 0$ and initializing $\B{b}$ as a spherical tensor results in an isotropic model.
 	\subsection{Thermodynamic consistency}\label{ssec:appendix_a:thermo}
 		To improve numerical stability in Eq.~\eqref{eq:damage_evol_0}, the Heaviside function $\mathcal{H}\of{f}$ and the Macaulay brackets $\mac{\dot{f}}$ are approximated by differentiable functions. The approximations read
 		\begin{alignat}{2}
 			\mac{\dot{f}} &\approx \frac{\dot{f}}{2} + \frac{\delta_\mr{M}}{2} \, 	\ln\of{\cosh\of{\frac{\dot{f}}{\delta_\mr{M}}}} + C_1 &&=: m\of{\dot{f}}\label{eq:approx_m}
 			\\
 			\mathcal{H}\of{f} &\approx \totDer{m\of{f}}{f} = \frac{1}{2} + \frac{1}{2} \, 	\tanh\of{\frac{f}{\delta_\mr{H}}} &&=: h\of{f}\label{eq:approx_h}
 		\end{alignat}
 		where the constant $C_1 = \tfrac{\ln\of{2}\,\delta_\mr{M}}{2}$ is chosen such that $m\of{\dot{f}} \geq 0 \quad \forall \quad \dot{f} \in \mathbb{R}$. Owing to this choice and the boundedness of the hyperbolic tangent, the functions in \cref{eq:approx_m,eq:approx_h} remain nonnegative. The numerical parameters are chosen as $\delta_\mr{M} = 10^{-4}$ (Eq.~\eqref{eq:approx_m}) and $\delta_\mr{H} = 10^{-2}$ (Eq.~\eqref{eq:approx_h}). This regularization can introduce a small approximation error that depends on the chosen values of $\delta_\mr{M}$ and $\delta_\mr{H}$. To quantify this effect, we repeated the simulations of Fig.~\ref{fig:calibration2}~(a) using smoothing-parameter sets varied over two orders of magnitude. For the case with a stress amplitude of $730\,\si{\mega\pascal}$, the predicted cycles to failure differed by approximately $0.1\,\%$ across the tested parameter range. The corresponding changes for the remaining investigated stress amplitudes were likewise negligible. We therefore conclude that, when the smoothing parameters are selected sufficiently small relative to the relevant scales of $f$ and $\dot{f}$, the residual numerical damage accumulation does not significantly affect the predicted fatigue life in the investigated high-cycle regime.
 		
 		The evolution equation of $\B{b}$ in its original form and its numerically smoothed counterpart read
 		\begin{align}\label{eq:dmgevol_final}
 			\dot{\B{b}} &= \mathcal{H}\of{f} \, \mac{\dot{f}} \, \partDer{\Gamma\of{\B{\beta}}}{\B{\beta}}, \qquad
 			\dot{\B{b}}_\text{smth} = h\of{f} \, m\of{\dot{f}} \, \partDer{\Gamma\of{\B{\beta}}}{\B{\beta}}
 		\end{align}
 		and the corresponding reduced mechanical dissipation inequalities follow
 		\begin{align}
 			\mathcal{D}^\text{red} &= \dot{\B{b}} : \B{\beta} = \overbrace{\mathcal{H}\of{f}}^{\text{$\geq 0$}} \, \overbrace{\mac{\dot{f}}}^{\text{$\geq 0$}} \,
 			\underbrace{\gamma\of{f}}_{\text{chosen $\geq 0$}}\,
 			\overbrace{\partDer{\Gamma\of{\B{\beta}}}{\B{\beta}} : \B{\beta}}^{\text{$\geq 0$ Eq.\eqref{eq:proof_gsm}}} \geq 0\, \text{,} \label{eq:dis_ineq}
 			\\
 			\mathcal{D}^\text{red}_\text{smth} &= \dot{\B{b}}_\text{smth} : \B{\beta} = \underbrace{h\of{f} \, m\of{\dot{f}}}_{\text{chosen $\geq 0$}} \,
 			\gamma\of{f}\,
 			\partDer{\Gamma\of{\B{\beta}}}{\B{\beta}} : \B{\beta} \geq 0 \label{eq:dis_ineqs}\, \text{.}
 		\end{align}
 		By choosing the scaling function~\eqref{eq:end_gamma} to be nonnegative for all real values of $f$ and the potential~\eqref{eq:end_Gamma_pot} to be convex, nonnegative, and to contain the origin, the dissipation inequality is satisfied. For the continuous approximation, the functions defined in \cref{eq:approx_h,eq:approx_m} must additionally remain nonnegative for all real values of $f$.
		
		The evolution direction derived from the potential~\eqref{eq:end_Gamma_pot} reads
		\begin{align}
			\partDer{\Gamma\of{\B{\beta}}}{\B{\beta}} &= \frac{\eta_\mr{i}\,\text{tr}\of{\B{\beta}}\,\B{I} + \eta_\mr{a}\,\B{\beta}}{\left[\eta_\mr{i}+\eta_\mr{a}\right] \, \Gamma\of{\B{\beta}}}
		\end{align}
		and its scalar product with $\B{\beta}$ appearing in the dissipation inequality~\eqref{eq:dis_ineq} and~\eqref{eq:dis_ineqs} reads
		\begin{align}\label{eq:proof_gsm}
			\partDer{\Gamma\of{\B{\beta}}}{\B{\beta}}:\B{\beta} &= \frac{\eta_\mr{i}\,\text{tr}\of{\B{\beta}}^2 + \eta_\mr{a}\,\text{tr}\of{\B{\beta}^2}}{\left[\eta_\mr{i}+\eta_\mr{a}\right] \, \Gamma\of{\B{\beta}}} = \frac{\Gamma\of{\B{\beta}}^2}{\Gamma\of{\B{\beta}}} = \Gamma\of{\B{\beta}} \geq 0 \quad \forall \quad \B{\beta} \in \text{Sym}(3)\,\text{.}
		\end{align}
		The energy-release-rate tensor (Eqs.~\eqref{eq:beta0}\textsubscript{(2)},~\eqref{eq:energyreleaseratefull_mcr},~\eqref{eq:energyreleaseratereduced_mcr}) is symmetric provided the strain measure is symmetric.
		
		\begin{remark}
			The thermodynamic constraint imposed on the scaling function~\eqref{eq:end_gamma} is nonnegativity. Choosing the scaling function to be additionally monotonically increasing ensures that damage evolution accelerates with increasing distance from the endurance surface.
		\end{remark}
	
		\begin{remark}
			Integrity recovery can occur in the present physically sound framework when enforcing thermodynamic consistency together with omitting the mixed term from~Eq.\eqref{eq:energyreleaseratefull_mcr} to Eq.~\eqref{eq:energyreleaseratereduced_mcr} by following~\cite{Ekh03} (mathematically, $\B{b}:\B{\ve}_{+} + \trace{\B{\ve}_{-}}$ can cause a change of sign). A careful numerical treatment in line with thermodynamic constraints is thus required. Although recovery has not been observed in the present computations, it was rigorously prevented by halting damage evolution during these states.
		\end{remark}
	\subsection{Direct gradient enhancement}\label{ssec:naive_enhancement}
		A natural way to incorporate the gradient enhancement within the endurance-surface framework is to employ the energy-release-rate tensor $\B{\beta}\enh$ (Eq.~\eqref{eq:duals_enh}) derived from the gradient-enhanced Helmholtz energy~\eqref{eq:helmholtz_enh}:
		\begin{align}\label{eq:endurance_surface_gr}
			\tilde{f}\enh\of{\B{\beta}\enh, \B{b}} &= \sqrt{\frac{\eta_\text{i}}{2} \, \left[ \B{b}^{r} : \B{\beta}\enh \right]^2 + \frac{\eta_\text{a}}{2} \, \B{b}^{r} : \left[ \B{\beta}\enh \cdot \B{b}^{r} \cdot \B{\beta}\enh \right]} - 1 \, \text{,}
			\\
			\tilde{\Gamma}\enh\of{\B{\beta}\enh} &= \sqrt{\frac{\eta_\text{i}\, \trace{\B{\beta}\enh}^2 + \eta_\text{a}\, \trace{\B{\beta}\enh^2}}{\eta_\text{i}+\eta_\text{a}}} \, \text{.}
		\end{align}
		Consequently, the gradient enhancement is incorporated through the microforce tensor $\B{\omega}$ via $\B{\beta}\enh$. This approach is thermodynamically consistent in the limiting case $\B{\varphi} \rightarrow \B{b}$. However, in the numerical setting the driving force $\B{\beta}\enh$ may exhibit positive eigenvalues if the auxiliary field $\B{\varphi}$ deviates sufficiently from the local integrity tensor $\B{b}$. As a consequence, the evolution of the integrity tensor $\B{b}$ becomes positive in the corresponding principal directions, leading to increasing integrity over time, i.e., decreasing damage over time. The modified formulation introduced in section~\ref{ssec:endurance_enhanced} addresses this unphysical effect to obtain meaningful results.
\section{Plain concrete -- Compact tension specimen under monotonic loading}\label{ssec:2d_conv}

		\paragraph{Setup}
		A compact tension (CT) specimen is discretized using approximately $\left\{1500, 3000, 7600, 23000\right\}$ quadrilateral elements for different mesh resolutions. The elements in the predicted damage zone are refined, whereas larger elements are used in the outer regions where damage evolution is not expected. Plane-strain conditions are assumed. A geometrical notch is introduced as a predetermined failure location. The specimen is loaded through circular holes. The midpoint of the lower hole is fixed, while the upper hole is loaded by an arc-length method. Nonlinear constraints are imposed to ensure the nodes located on the hole boundaries maintain a constant distance of $r=40\,\si{\milli\meter}$ from the respective hole center. The MCR effect is disabled for the present setup.
		
\begin{figure}[ht]
			\centering
			\subfigreset
			\subbfigure[Geometry and boundary conditions of the CT specimen and force--displacement curves for the different mesh refinements under monotonic loading. Lengths are given in $\si{\milli\meter}$.]{%
				\psfrag{disp}[c][c]{\scalebox{0.8}{\makebox{displacement $u \, [\si{\milli\meter}]$}}}
				\psfrag{frc}[c][c]{\scalebox{0.8}{\makebox{force $F \, [\si{\newton}]$}}}
				\psfrag{fu}[l][c]{\scalebox{0.7}{\makebox{$u$,$F$}}}
				\psfrag{l1}[c][c]{\scalebox{0.4}{\makebox{$100$}}}
				\psfrag{l2}[c][c]{\scalebox{0.4}{\makebox{$180$}}}
				\psfrag{l3}[c][c]{\scalebox{0.4}{\makebox{$500$}}}
				\psfrag{l4}[c][c]{\scalebox{0.4}{\makebox{$130$}}}
				\psfrag{l5}[c][c]{\scalebox{0.4}{\makebox{$240$}}}
				\psfrag{l6}[c][c]{\scalebox{0.4}{\makebox{$20$}}}
				\psfrag{dm}[c][c]{\scalebox{0.4}{\makebox{$\diameter 80$}}}
				\vspace{0.3cm}
				\begin{tikzpicture} 
					\node[anchor=south west,inner sep=0] (main) at (0,0) {\includegraphics[width=0.48\linewidth]{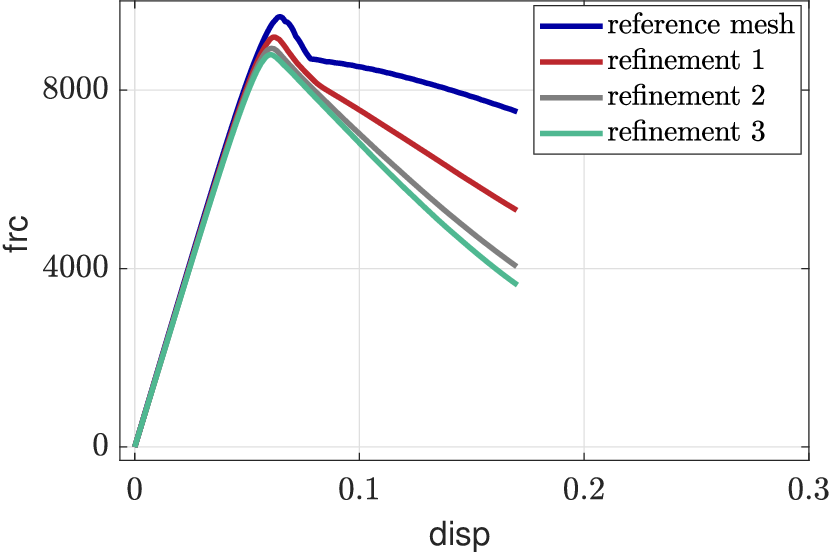}};
					\begin{scope}[x={(main.south east)},y={(main.north west)}]
						\node[anchor=north east] at (0.96,0.74) {\includegraphics[width=0.15\linewidth]{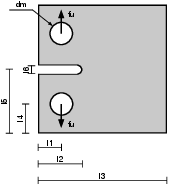}};
					\end{scope}
			\end{tikzpicture}%
			}
			\hspace{0.1cm}
			\subbfigure[Integrity fields of the different mesh refinements at $u=0.17\,\si{\milli\meter}$.]{\psfrag{u}[l][c]{\scalebox{0.8}{\makebox{1.0}}}
				\psfrag{l}[l][c]{\scalebox{0.8}{\makebox{0.0}}}
				\psfrag{t}[c][c]{\rotatebox{90}{\scalebox{0.8}{\makebox{integrity $b$}}}}
				\raisebox{0.6cm}{\begin{tikzpicture}
					\node[anchor=south west,inner sep=0] (main) at (0,0) {\includegraphics[width=0.3\linewidth]{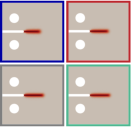}};
					\begin{scope}[x={(main.south east)},y={(main.north west)}]
						\node[anchor=north east] at (-0.1,0.7) {\includegraphics[width=0.025\linewidth]{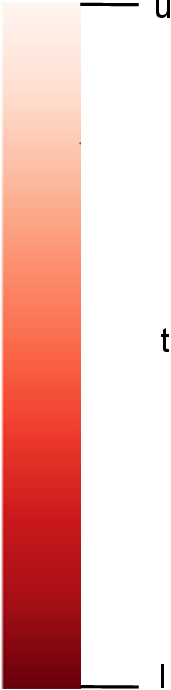}};
					\end{scope}
			\end{tikzpicture}}
			}
			\caption{Plain concrete -- Compact tension specimen under monotonic loading: geometry, boundary conditions, structural response and damage fields under monotonic loading of four discretizations (\{blue: $1500$, red: $3000$, gray: $7600$, cyan: $23000$\} elements).}\label{fig:ct2d_setup_fu}
		\end{figure}

		\paragraph{Discussion}
		The results are summarized in Fig.~\ref{fig:ct2d_setup_fu}.
		The macroscopic responses for the different mesh discretizations are shown in Fig.~\ref{fig:ct2d_setup_fu}~(a). After the initial elastic regime, all curves reach a peak load followed by softening caused by the progressive degradation. While the coarsest mesh exhibits clear mesh-dependent artefacts in the force--displacement curve, successive refinements lead to mesh convergence. The corresponding damage fields are shown in Fig.~\ref{fig:ct2d_setup_fu}~(b) for a displacement of $u = 0.17\,\si{\milli\meter}$. They reveal a well-defined crack path propagating through the center of the specimen.
		
\section*{Acknowledgements}
Financial support from the German Research Foundation (DFG) via SFB/TRR 188 (project number 278868966), project C01, is gratefully acknowledged. We also gratefully acknowledge the computing time provided on the Linux HPC cluster at TU Dortmund University (LiDO3), partially funded in the course of the Large-Scale Equipment Initiative by the German Research Foundation (DFG) (project number 271512359).

{\small
	\bibliographystyle{elsarticle-num}
	\bibliography{references_main.bib}
}

\end{document}